\documentclass{emulateapj}
\usepackage{rotating,graphicx}
\def\plotfiddle#1#2#3#4#5#6#7{\centering \leavevmode
    \vbox to#2{\rule{0pt}{#2}}
    \includegraphics{#1}}
\def\alwaysmath#1{\ifmmode{#1}\else{$#1$}\fi}

\def\Msun{M_{\odot}}
\def\Mvir{M_{\rm vir}}
\def\Rvir{R_{\rm vir}}
\def\cvir{c}
\def\Vvir{V_{\rm vir}}
\def\Vmax{V_{\rm max}}
\def\dvir{\Delta_{\rm vir}}

\def\kms{ {\rm km}\,{\rm s}^{-1}}
\def\beq{\begin{equation}}
\def\eeq{\end{equation}}
\def\beqa{\begin{eqnarray}}
\def\eeqa{\end{eqnarray}}
\def\lsim{\lower0.6ex\vbox{\hbox{$ \buildrel{\textstyle <}\over{\sim}\ $}}}
\def\gsim{\lower0.6ex\vbox{\hbox{$ \buildrel{\textstyle >}\over{\sim}\ $}}}
\def\Lsun{L_{\odot}}

\begin{document}

\title{Tracing Galaxy Formation with Stellar Halos I: Methods }
\author{James  S.        Bullock\altaffilmark{1} \&
Kathryn V. Johnston \altaffilmark{2}}
\altaffiltext{1}{Department of Physics \& Astronomy,
        University of California, Irvine, CA 92697, USA; bullock@uci.edu}
\altaffiltext{2}{Van Vleck Observatory, Wesleyan University, Middletown, CT 06459, 
USA; kvj@astro.wesleyan.edu}

\begin{abstract}
If  the favored hierarchical cosmological model   is correct, then the
Milky Way  system  should  have  accreted $\sim 100-200$ luminous satellite
galaxies in the past $\sim 12$ Gyr.
 We model this  process using a
hybrid   semi-analytic  plus   N-body   approach  which  distinguishes
explicitly between the evolution of light and  dark matter in accreted
satellites.  This distinction  is essential to our ability to produce a  realistic
stellar halo, with mass and density profile much  like that of our own
Galaxy, and a surviving satellite population that 
matches  the    observed  number   counts and    structural  parameter
distributions of the satellite galaxies of the Milky Way.
Our model stellar halos have  density profiles which typically drop
off with radius faster than those of the dark matter. They
are assembled  from the inside out, with the majority of mass ($\sim  80\%$)
coming from the  $\sim  15$ most massive  accretion  events.
The satellites that contribute
to the stellar halo have median accretion times of
$\sim 9$ Gyr in the past, while surviving satellite systems have median accretion
times of $\sim 5$ Gyr in the past.
This implies that  stars associated with  the inner  halo should  be  quite
different chemically from  stars   in surviving satellites  and   also
from stars  in the outer halo  or those liberated  in recent
disruption events.
We briefly discuss the expected spatial structure
and phase space structure for halos formed in this manner.
Searches for this type of structure
offer a direct  test  of whether cosmology  is
indeed hierarchical on small scales.
\end{abstract}

\section{Introduction}

There has been a  long tradition of searching in  the stellar  halo of
our Galaxy for signatures of its formation.  Stars in the halo provide
an important avenue for testing  theories of galaxy formation
because  they have long orbital time
periods, have likely suffered  little  from dissipation effects, and tend   to
inhabit the  outer regions   of   the Galaxy where the    potential is
relatively smooth  and slowly  evolving.  The  currently  favored Dark
Energy + Cold Dark  Matter ($\Lambda$CDM) model of structure formation
makes the  specific prediction that  galaxies like  the Milky Way form
hierarchically, from a series of  accretion events involving lower-mass
systems.   This  leads naturally  to the  expectation that the stellar
halo should be formed primarily from  disrupted, accreted systems.  In
this work, we  develop an explicit, cosmologically-motivated model for
stellar  halo formation  using    a hybrid N-body plus   semi-analytic
approach.  Set within the  context of $\Lambda$CDM,  we use this model
to test the general consistency of the hierarchical formation scenario
for  the stellar halo and  to provide predictions for upcoming surveys
aimed  at probing  the accretion history  of the  Milky Way and nearby
galaxies.

In  a   classic study,  \citet{els62} used proper   motions and radial
velocities  of  221 dwarfs to  show  that those with lower metallicity
(i.e. halo stars) tended to move on more highly eccentric orbits. They
interpreted  this trend  as  a  signature of   formation of the  lower
metallicity  stars   during a rapid  radial  collapse.    In contrast,
\citet{sz78} suggested that the wide range of  metallicites found in a
sample of  19 globular clusters at  a variety  of Galactocentric radii
instead     indicated   that the    Galaxy   formed  from  the gradual
agglomeration of many sub-galactic sized pieces.  A recent analysis of
1203 metal-poor Solar  neighborhood stars, selected without  kinematic
bias  \citep{chiba00}, points to the truth   being some combination of
these  two  pictures: this  sample  contained a small concentration of
very low metallicity stars  on highly eccentric orbits (reminiscent of
Eggen, Lynden-Bell \&  Sandage's, 1962 work)  but  otherwise showed no
correlation  of  increasing   orbital   eccentricity  with  decreasing
metallicity.

In the  last decade, much more direct  evidence for the lumpy build-up
of the  Galaxy  has  emerged  in the   form of   clumps of  stars   in
phase-space (and, in  some cases, metallicity) both  relatively nearby
\citep{majewski96,helmi99b}  and at   much larger  distances. The most
striking  example  in the   latter category is    the discovery of the
Sagittarius dwarf galaxy \citep{ibata94,ibata95} --- hereafter Sgr ---
and its associated trails of debris
\citep[see][for an overview of the many detections]{majewski03}
which    have    now  been  traced     entirely   around   the  Galaxy
\citep{ibata01b,majewski03}. Large scale  surveys of the  stellar halo
are                         now                               underway
\citep{majewski00,morrison00,yanny00,ivezic00,newberg02},  and    have
uncovered  additional    structures,    not  associated    with    Sgr
\citep{newberg03,martin04,rocha-pinto04}.  Moreover, recent  advances  in instrumentation are
now  permitting searches for  and  discoveries of analogous structures
around other galaxies in the form of overdensities in integrated light
\citep{shang98,zheng99,forbes03,pohlen04} or, in the case of M31, star
counts  \citep{ibata01a,ferguson02,zucker04}.  Given  this plethora of
discoveries,  there  can   be little  doubt  that   the   accretion of
satellites has been  an important contributor to  the formation of our
and   other       stellar halos.  In    addition,    both  theoretical
\citep{abadi03,brook05a,robertson05} and observational
\citep{gilmore02,yanny03,ibata03,crane03,rocha-pinto03,frinchaboy04,helmi05}
work is  beginning  to   suggest that  some  significant  fraction  of
the Galactic disk could also have been formed this way.

All  of the  above discoveries  are in  qualitative agreement with the
expectations of hierarchical structure formation
\citep{peebles:65,ps74,bfpr:84}.
As the prevailing variant of  this picture, $\Lambda$CDM is remarkably
successful at reproducing a wide  range of observations, especially on
large scales
\citep[e.g.][]{eisenstein:05,maller:05,tegmark:04,spergel:03,percival:02}.
On  sub-galactic scales, however, the   agreement  between theory  and
observation            is             not           as         obvious
\citep[e.g.][]{simon05,kazantzidis04b,DB:04}.   
Indeed, the  problems   explaining
galaxy rotation curve data, dwarf galaxy counts, and galaxy disk sizes
have lead some to   suggest modifications  to the  standard  paradigm,
including an  allowance  for  warm  dark matter  \citep[e.g.][]{SL04},
early-decaying dark  matter \citep{kaplinghat:05},  or       non-standard
inflation  \citep{zb:02,zb:03}.  These modifications generally suppress
fluctuation amplitudes on small scales, driving sub-galactic structure
formation towards a more monolithic, non-hierarchical collapse.  These
issues  bring to  sharper focus  a  fundamental question
in cosmology today: is structure
formation truly hierarchical  on small  scales?  Stellar halo  surveys
offer powerful data sets for directly answering this question.

Numerical simulations of individual satellites disrupting about parent
galaxies can  in many  cases  provide  convincing similarities to  the
observed phase-space lumps.  These models allow the observations to be
interpreted in terms of the mass and orbit of the progenitor satellite
\citep[e.g.,][]{velazquez95,johnston95,johnston99b,johnston99c,helmi01,helmi03b,law05},
and even the potential of the galaxy in which it is orbiting
\citep{johnston99a,murali99,ibata01b,ibata04,johnston05}.
Nevertheless, a true test of  hierarchical  galaxy formation will require
robust predictions for   the frequency and  character  of the expected
phase space structure of the halo.

Going  beyond  qualitative statements  to  model the full stellar halo
(including substructure) within    a  cosmological context
is non-trivial.  The largest contributor of  substructure to our own halo
is  Sgr, estimated to  have a  currently-bound  mass of order $3\times
10^8 M_\odot$ \citep{law05}.  Even the highest resolution  cosmological
N-body simulations would not  resolve such an object  with more than a
few hundred particles, which  would permit only a poor  representation
of the phase-space structure of its  debris \citep[see][for an example
of   what  can  currently  be done   in   this  field]{helmi03a}. Such
simulations are computationally  intensive,  so the cost of  examining
more than a handful  of halos  is prohibitive  and it is  difficult to
make  statements about the variance  of properties of halos that might
be seen in  a  large sample  of galaxies. Moreover,  such  simulations
in general only follow  the dark matter component of  each galaxy not the stellar
component.   In their studies of thick disk and inner halo
formation, Brook and collaborators \citep{brook03,brook04a,brook04b,brook05a,brook05b}
have modeled the stellar components directly  by simulating the evolution of individual galaxies as
isolated spheres of dark matter and gas with small-scale
 density fluctuations superimposed to account for
the large-scale cosmology.
However, their sample size remains small and, 
though they are able to make general statements about the properties of their stellar halos,
their resolution would prohibit a detailed phase-space analysis.

An alternative  is to take  an analytic  or semi-analytic approach  to
halo    building \citep[e.g.][]{bullock01,johnston01,taylor04}. 
This allows the
production of many halos, and the potential of including prescriptions
to follow  the stars separately   from the dark matter.  However, such
techniques use only  approximate descriptions of  the dynamics and are
unable to follow   the  fine  details  of  the  phase-space  structure
accurately.

In this study we develop a hybrid scheme, which draws on the strengths
of  each of   the former techniques  to  build  high resolution,  full
phase-space models of a statistical sample of {\it stellar} halos. Our
approach is  to  vastly  decrease the computational   cost  of a  full
cosmological simulation by modeling  only those accretion events  that
contribute  directly   to  the stellar   halo   in detail  with N-body
simulations,  and  to   represent   the  rest of   the    galaxy  with
smoothly-evolving analytic functions.  The  baryonic component of each
contributing event is followed using semi-analytic prescriptions.

The purpose of this paper is to describe our method, its strengths and
limitations   (\S  2), present   the  results  of   tests of  the
consistency of our   models with general   properties of galaxies  and
their satellite systems (\S 3) and outline some
implications (\S 4).
We summarize the conclusions in \S 5. 
In further  work we will go on
to compare  the full phase-space structure  of our halos in  detail to
observations and to examine the evolution of dark and light matter
in satellite galaxies after their accretion.

\section{Methods}

Our  methods can  be  broadly separated into:   (I) a {\it simulation}
phase, which follows the phase-space evolution of the dark matter; and
(II) a {\it   prescription} phase, which  embeds   a stellar mass with each
dark matter particle.  Specifically:

\bigskip
\noindent{Phase I: Simulations} 

	\begin{enumerate}
		\item We generate merger  trees for our parent galaxies using
		the method outlined  in \citet{somerville99} based  on
		the Extended-Press-Schechter (EPS)   
		formalism  \citep[][---  see \S
		2.1)]{lc93}. 

		\item  For each event in step IA, we run
		a high-resolution N-body  simulation that
	        tracks the evolution of the dark matter component of a satellite 
	        disrupting  within an analytic, time-dependent,
		parent galaxy + host halo potential (see
		\S 2.2). 	

	\end{enumerate}


\noindent Phase II: Prescriptions


\begin{enumerate}

		\item We follow the gas accretion history of each 
		satellite prior to falling into the parent and
		track its star-formation  rate using
		cosmologically-motivated, semi-analytic prescriptions (see
		\S 2.3).


		\item We embed the stellar components generated in
		step IIA within each dark matter satellite by assigning a  variable mass-to-light ratio to
		every particle that is tracked in the (Phase I)
		N-body simulations (see \S 2.4).

	\end{enumerate}


We  consider the two-phase approach a necessary and
acceptable  simplification  since    it   allows  us     to   separate
well-understood  and   justified  approximations  in  Phase   I   from
prescriptions that can be adjusted  and refined during  Phase II.   In
addition, this separation allows us to save computational time and use
just  one set of  dark matter  simulations  to  explore the  effect of
varying the details of how baryons are assigned  to each satellite.  A
more   complete  discussion of the strengths    and limitations of our
scheme is given in \S2.5.

\subsection{Cosmological Framework}

Throughout this work we assume a $\Lambda$CDM cosmology with
$\Omega_{\rm m}=0.3$,  $\Omega_{\rm  \Lambda}=0.7$,     $\Omega_{\rm
b}h^2=0.024$, $h=0.7$,  and $\sigma_8 = 0.9$.  The implied
baryon fraction is $\Omega_b/\Omega_{\rm m} = 0.16$. 

We focus on the formation of stellar halos for ``Milky-Way'' type
galaxies.  In  all cases our $z=0$ host  dark matter halos have virial
masses $M_{\rm  vir,0}=1.4  \times 10^{12}  M_\odot$, corresponding
virial radii $R_{\rm vir,0}=282$  kpc, 
and virial velocities $\Vvir = 144 \kms$.
The quantities $\Mvir$ and  $\Rvir$ are related by
\beq
\label{eq:Rvir}
\Mvir = \frac{4\pi}{3} \rho_{\rm M}(z) \dvir(z) \Rvir^3,
\eeq
where $\rho_{\rm M}$ is the average matter density of the Universe and
$\Delta_{\rm  vir}$ is the  ``virial  overdensity''.  In the cosmology
considered here, $\dvir(z=0)  \simeq 337$, and $\dvir \rightarrow 178$
at $z \gsim 1$ \citep{bn:88}.   The virial velocity is defined as
$\Vvir \equiv \sqrt{G \Mvir/\Rvir}$.  

We generate a total of eleven random realizations of stellar halos.
General properties of all eleven 
are summarized in Table \ref{halo_tab}.
Any variations in our results for stellar halos among these are
determined by   differences in their accretion   histories.
In all subsequent figures we present results for four stellar
halos (1,2,6, and 9) chosen to span the  range of properties seen in our full sample.

\subsubsection{Semi-analytic accretion histories}

\begin{figure}
\plotfiddle{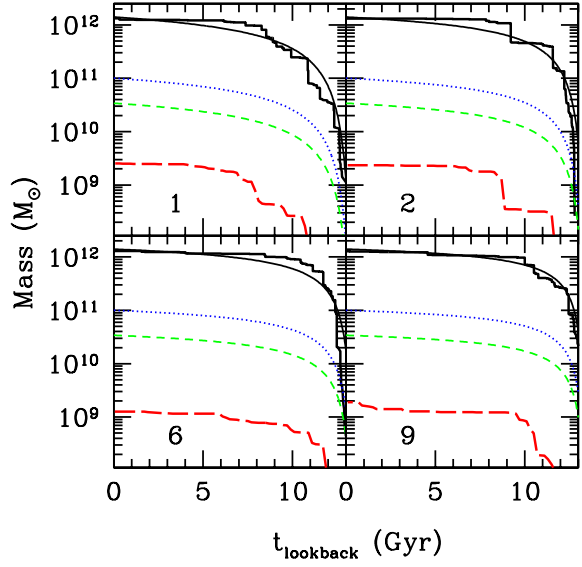}{3.5in}{0}{40}{40}{-150}{-50}
\caption{The mass assembly histories for halos 1, 2, 6, and 9.  
Solid jagged lines show the histories for the dark matter halos,
$\Mvir(<t)$, generated via the  EPS merger tree method.   Smooth solid
lines   show our best-fit ``smooth''    accretion history used for the
background    potential    in  our     N-body   simulations  (Equation
\ref{growth}). Smooth dotted and short-dashed  lines  show the evolution in
our  disk   and bulge  component masses   used   to set   the galactic
contributions   to the  N-body  potentials  (Equations \ref{diskeqn} -
\ref{bulgeqn}). 
The long-dashed lines show our (main) results for stellar halo
assembly histories.    Stellar mass is  assigned  to  the stellar halo
component  at  the time it becomes  unbound  from an  accreted satellite
halo.  Note that the stellar halo in realization 6 is relatively 
small compared to the other systems whose dark matter halos are all
the same size.  This is because dark matter halo 6 accumulated
most of its mass relatively
early, when the stellar mass fractions of accreted satellites were small.
Note also that halo 9 has experienced a very recent, massive disruption
 accretion event, which causes its stellar halo mass to increase sharply
at a recent lookback time.  This event is seen clearly in the image of
halo 9 shown in Figure \ref{haloviz2_fig}.  
\label{buildup_fig}}
\end{figure}

We track the mass accretion and satellite acquisition of each parent
galaxy by constructing merger trees using the statistical
 Monte Carlo  method of \cite{somerville99} based on the EPS formalism
\citep{lc93}.  This method gives us a record of the the masses
and accretion times   of all satellite  halos and hence allows us
to follow the mass accretion history  of each parent
as a function of lookback time.   We explicitly note  all
satellites more massive than $M_{\rm min} = 5 \times 10^6 \Msun$ 
 and treat   all smaller accretion events as   diffuse
mass accretion.  Column 2 of Table \ref{halo_tab} lists the total number of such events for each
simulated halo. For further details see
\citet{lc93,somerville99,zb:03}.  Four examples of the cumulative mass 
accretion histories of parent galaxies generated in this manner are shown by the (jagged)
solid lines in Figure \ref{buildup_fig}.

\subsubsection{Satellite orbits} 

Upon  accretion onto  the host,  each satellite   is assigned  an initial
orbital  energy based  on  the range  of  binding energies observed in
cosmological simulations \citep[][]{klypin99}.
This is done by placing  each satellite on
an initial orbit of energy equal to the  energy of a circular orbit of
radius $R_{\rm circ} = \eta \Rvir$, with $\eta$ drawn randomly
from a uniform  distribution  on the interval $[0.4,0.8]$.
Here  $\Rvir$ is the virial radius
of the host halo at the time of accretion.   We assign
each subhalo an initial specific angular momentum $J = \epsilon
J_{\rm circ}$,  where  $J_{\rm circ}$  is the specific   angular   momentum of the
aforementioned   circular orbit and  $\epsilon$  is  the {\em  orbital
circularity}, which takes a value  between $0$  and $1$.  We choose
$\epsilon$     from the binned distribution    shown   in Figure 2  of
\cite{zb:03},  which  was designed  to   match the cosmological N-body
simulation  results of  \cite{ghigna98},  and    is  similar  to   the
circularity  distributions found    in  more recent  N-body   analyses
\citep{zentner04,benson05}.  Finally, the plane of the orbit is drawn 
from a uniform distribution covering the halo sphere.

\subsubsection{Dark matter density distributions}
\label{sec:dm}

We model all satellite and parent halos with the spherically 
averaged density profile of \citet{nfw96} (NFW):
\beq
\label{eq:nfw_profile}
\rho_{_{\rm NFW}(r)} = \rho_{\mathrm{s}} \left(\frac{r}{r_{\rm halo}}\right)^{-1}
\left(1 + \frac{r}{r_{\rm halo}} \right)^{-2} ,
\eeq
where   $r_{\rm halo}$ ($\equiv r_{s}$ in   NFW) is the characteristic
inner   scale     radius    of    the   halo.     The   normalization,
$\rho_{\mathrm{s}}$, is set by the  requirement that the mass interior
to $\Rvir$   be equal  to  $\Mvir$.  The  value   of $r_{\rm halo}$ is
usually  characterized in   terms of  of  the  halo  ``concentration''
parameter: $\cvir  \equiv  \Rvir/r_{\rm halo}$.   The  implied maximum
circular  velocity for this  profile occurs  at  a radius $r_{\rm max}
\simeq 2.15 r_{\rm halo}$ and takes the value
$\Vmax  \simeq 0.466 \Vvir F(c)$,  where
$F(c) = \sqrt{c/[\ln(1+c)-c/(1+c)]}$.

For  satellites,  we set the value  of $\cvir$ using the
simulation results  of \citet{b01} and the  corresponding relationship
between  halo  mass, redshift, and  concentration  summarized by their
analytic model.   The median $\cvir$ relation for halos of mass $\Mvir$
at redshift $z$ is
given {\em approximately} by
\beq
	\cvir \simeq 9.6 \left(\frac{\Mvir}{10^{13} \Msun} \right)^{-0.13} (1+z)^{-1},
\eeq
although in practice we use the full analytic model discussed in \cite{b01}.

For   parent halos,  we     allow   their  concentrations to     evolve
self-consistently as their virial masses  increase, as has been seen  in
the N-body     simulations of  \citet{wechsler02}.  
Rather than represent the halo growth as a series of discrete accretion events,
we smooth over the Monte Carlo EPS merger tree by   
fitting the following functional form to the Monte Carlo  mass
accretion history for each halo:
\begin{eqnarray}
	M_{\rm vir}(a) & = & M_{\rm vir}(a_0) \exp \left[-2 a_c \left(\frac{a_0}{a} - 1 \right) \right].
\label{growth}
\end{eqnarray}
Here $a \equiv (1+z)^{-1}$ is the expansion factor, 
and $a_c$ is the fitting parameter, corresponding
to the value of the expansion factor at
a characteristic ``epoch of collapse''. 
\citet{wechsler02} demonstrated that
the value of $a_c$ connects in a one-to-one fashion with the
halo concentration parameter \citep{wechsler02}:
\beq
	c(a) = 5.1 {a \over a_c}.
\label{cofz}
\eeq
Halos that form earlier (smaller $a_c$'s) are more concentrated.

Example fits to four of our halo mass accretion histories are shown by
the  smooth  solid  lines  in  Figure   1. 
The $a_c$ values for each of the halos
in this analysis are listed in the third column of Table 1.
Typical host halos in our sample
have $\cvir \simeq 14$ at $z=0$, scale radii $r_{\rm halo}
\simeq 20$kpc, and maximum circular velocities $\Vmax \simeq 190 \kms$.

\subsection{N-body simulations of dark matter evolution}

Having determined the mass, accretion time and orbit of each satellite (\S2.1.1 and \S2.1.2), 
and the evolution  the potential into which it is falling (\S2.1.3), we next run 
individual N-body simulations
 to track  the dynamical evolution  of each satellite halo separately.
 We follow only those that contain a  significant  stellar
component (see \S 2.3 below).  In  practice, this  restricts  our analysis  to  satellite
halos  more  massive  than  $\Mvir \gsim 10^8   \Msun$ --- the number of such satellites 
infalling into 
each parent is listed in column 5 of Table \ref{halo_tab}.   Based on our
star-formation prescription discussed in  \S 2.3, systems smaller than
this  never contain an  appreciable number  of   stars and thus  don't
contribute significantly to the stellar halo.

\subsubsection{The parent galaxy potential}

The parent galaxy is  represented by a three-component bulge/disk/dark
halo potential which we allow to evolve with time as the halo accretes
mass.  The (spherically-symmetric) dark  halo potential at each epoch $a$
is given by the NFW potential generated by the dark matter distribution in equation 
(\ref{eq:nfw_profile})
\beqa
	\Phi_{\rm halo}(r)  =  -{G M_{\rm halo}\over r_{\rm halo}}{1 \over (r/r_{\rm halo})} \ln\left({r \over r_{\rm halo}}+1\right),
\label{haloeqn}
\eeqa{equation}
where $M_{\rm halo}= M_{\rm halo}(a)$  and $r_{\rm halo} = r_{\rm halo}(a)$ 
are the instantaneous mass and length scales of the halo respectively. 
The halo mass scale is  related to the virial mass via
\begin{eqnarray}
	M_{\rm halo}&=&{M_{\rm vir} \over \ln (c+1) -c/(c+1)}. 
\end{eqnarray}

The disk and bulge are assumed to grow in mass and scale with the halo virial mass and radius:
\begin{equation}
        \Phi_{\rm disk}(R,Z)=-{GM_{\rm disk} \over
                 \sqrt{R^{2}+\left(R_{\rm disk}+\sqrt{Z^{2}+Z_{\rm disk}^{2}}\right)^{2}}},
\label{diskeqn}
\end{equation}
\begin{equation}
        \Phi_{\rm sphere}(r)=-{GM_{\rm sphere} \over r+r_{\rm sphere}},
\label{bulgeqn}
\end{equation}
where $M_{\rm disk}(a)=1.0  \times 10^{11} (M_{\rm vir}/M_{\rm vir,0})
M_{\odot}$,  $M_{\rm sphere}(a)=3.4 \times 10^{10} (M_{\rm vir}/M_{\rm
vir,0}) M_{\odot}$,  $R_{\rm disk}=6.5 (r_{\rm  vir}/r_{\rm vir,0})$ kpc, $Z_{\rm disk}=0.26
(r_{\rm   vir}/r_{\rm   vir,0})$ kpc  and   $r_{\rm sphere}=0.7(r_{\rm  vir}/r_{\rm
vir,0})$ kpc.

\subsubsection{Satellite initial conditions}

\begin{figure}
\plotfiddle{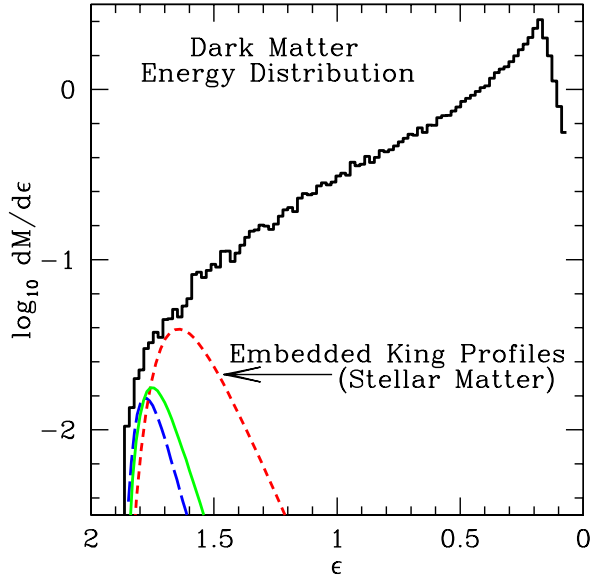}{3.5in}{0}{40}{40}{-150}{-50}
\caption{The energy distribution function of our initial condition
dark matter halo (histogram, $dM/d\epsilon$) along with three example energy distributions for
stellar matter,  $(dM/d\epsilon)_\star$, in satellites.  
The mass to light ratio of each particle of energy $\epsilon$ is assigned
based on the ratio of $(dM/d\epsilon)_\star$ to $(dM/d\epsilon)$.
Energy in this plot is in units of $G M_{35}^2/2R_{\rm halo}$.
\label{embed_fig}}
\end{figure}

We use $10^5$ particles  to represent the dark matter in each accreted satellite.
Particles are initially distributed
as an isotropic NFW model, with mass and scale  chosen as described in
\S2.1.2.  The  phase-space   distribution  function   is   derived  by
integrating over the density and potential distributions
\begin{equation}
	f(\epsilon)= {1 \over 8\pi^2} \left[ 
\int_0^\epsilon {d^2\rho \over d\Psi^2} {d\Psi \over \sqrt{\epsilon -\Psi}} + \frac{1}{\sqrt{\epsilon}} \left(\frac{d\rho}{d \Psi}\right)_{\Psi=0} \right].
\label{fofe_eq}
\end{equation}
with $\rho = \rho_{_{\rm NFW}}$ and where
$\Psi = -\Phi_{_{\rm NFW}} + \Phi_0$
  is the  relative
potential (such that $\Psi \rightarrow  0$ as  $r  \rightarrow   \infty$) 
and $\epsilon = \Psi - v^2/2$ is the relative energy
\citep[see][for discussion]{bt87}.  This distribution function
 is used (in tabulated form) to generate a random realization. This
ensures  a  stable   satellite  configuration  --- initial  conditions
generated by instead assuming a local Maxwellian velocity distribution
have been shown to evolve \citep{kazantzidis04a}.  Given
$f(\epsilon)$, the
differential energy distribution follows in a straightforward
manner from the density of states, $g(\epsilon)$,
\beq
\frac{d M}{d \epsilon} = f(\epsilon)g(\epsilon), \quad 
g(\epsilon) \equiv 16 \pi^2 \int_0^{r_\epsilon} \sqrt{2(\Psi - \epsilon)} r^2 dr,
\eeq
where $r_\epsilon$ is the largest energy that can be reached by a star of relative energy $\epsilon$.  The differential energy distribution
for our initial halo is shown
by the solid histogram in Figure \ref{embed_fig}.  We see that the
majority of the (dark matter) material in an infalling satellite
is quite loosely bound.

Rather    than generating  a    unique $f(\epsilon)$  and
particle distribution for each  satellite in each accretion history, a
single initial conditions file  with  unit mass  and scale,  and outer
radius   $R_{\rm out}  =  35  r_{\rm   halo}$ ($ = 35$ in our units)    
is  used for all
simulations with  masses and  scales appropriately rescaled  for each
run.   
Since all of our accreted satellites have concentrations $c < 35$, our
set up effectively  allows  each accreted satellite's mass  profile to
extend beyond its virial radius for several  scale lengths.  We do not
expect this simplification to significantly affect our results because
the the light matter is always embedded at the very central regions of
the halo  ($r_\star \lsim  r_{\rm halo}$)  and the outer  material  is
always quickly stripped  away from the outer parts  of  the halos upon
accretion.  

In \S 2.4 we discuss our method  of ``embedding'' star particles within
the cores  the  accreted  satellite  dark  halos.

\subsubsection{Satellite evolution}

The  mutual interactions  of the  satellite   particles are calculated
using a basis function expansion code \citep{hernquist92}. The initial
conditions file for the satellite is allowed to relax in isolation for
ten  dynamical times using  this  code to  confirm stability. For each
accretion event a single simulation is run, following the
evolution of the relaxed satellite under the  influence of its own and
the parent galaxy's potential, for the  time since it was accreted (as
generated by  methods in  \S 2.1.1)  along the  orbit chosen at random
from the distribution discussed in \S2.1.2.
\citep[Note that simulations of satellite accretions in static NFW potentials using this code produced results identical to those reported in][]
{hayashi03}

Using this approach, the satellites are  not influenced by each other,
other than   through the     smooth  growth  of  the  parent    galaxy
potential.   Nor does  the  parent  galaxy   react   to the  satellite
directly. In order to mimic the expected decay of the satellite orbits
due to dynamical  friction (i.e. the interaction  with the parent), we
include a  drag term on  all particles within  two tidal radii $r_{\rm
tide}$  of the   satellite's   center,   of  the  form     proposed by
\citet{hashimoto03} and modified for NFW hosts by \citet{zb:03}.  
This   approach includes   a   slight  modification  to   the standard
Chandrasekhar    dynamical  friction formula \citep[e.g.][]{bt87}. The
tidal radius $r_{\rm tide}$ is calculated from the instantaneous bound
mass of the satellite $m_{\rm sat}$, the distance $r$ of the satellite
to the center of the  parent galaxy and  the mass of the parent galaxy
within that radius $M_r$ as $r_{\rm tide}=r (m_{\rm sat}/M_r)^{1/3}$.


\subsubsection{Increasing phase-space resolution with test particles}
\label{sec:test}

In this study,  we are most interested   in following the  phase-space
evolution of the stellar material associated with each satellite. This
is assumed to  be embedded deep within  each dark matter  halo (see \S
2.4) --- typically  only of order  $10^4$ of  the N-body particles  in
each satellite have any light associated with them at all. In order to
increase the statistical  accuracy  our analysis  we  sample the inner
12\% of  the energy distribution  with  an additional $1.2\times 10^5$
{\it  test} particles.  This does  not  increase the dynamic range our
simulation, but does  allow us to more finely  resolve the low surface
brightness features we  are interested in  with only a modest increase
in computational cost: we  gain a factor of  10 in particle resolution
with an increase of $\sim$25\% in computing time.
In this paper, we have used test particles only in 
generating the images shown in Figures \ref{haloviz1_fig} - \ref{phase_diagram2_fig}.

\subsection{Following the satellites' baryonic component}

We follow each satellite's baryonic component using the 
expected mass accretion history of each satellite halo (prior to falling into
the parent galaxy)
in order to  track the inflow of gas.   
The gas mass  is then  used to
determine the instantaneous star    formation rate and to    track the
buildup of stars within each halo.
The physics of galaxy formation is poorly understood,
and any attempt  to model star formation and gas inflow into galaxies
(whether semi-analytic or hydrodynamic)
necessarily require free parameters.
Our own prescription requires
three "free" parameters:
$z_{\rm re}$, the redshift of reionization (see \S \ref{sec:reion});
$f_{\rm gas}$, the fraction of baryonic material
in the form of cold gas (i.e. capable of forming stars) that remains bound to 
each satellite at accretion (see \S \ref{sec:gas});
and $t_\star$, the globally-averaged star formation
timescale (see \S \ref{sec:stars}). 

In the following subsections we describe how these parameters enter into our prescriptions,
and choose a value of $f_{\rm gas}$ consistent with observations.
In \S 3 we go on to demonstrate that   the     observed
characteristics of the stellar   halo  (e.g.   its mass,  and   radial
profile) and the Milky Way's satellite system (e.g. their number
and distribution in structural parameters)
provide strong  constraints   on the remaining free parameters and hence
the efficiency  of   star
formation  in low-mass   dark matter  halos in general.  

\subsubsection{Reionization}
\label{sec:reion}

Any  attempt to   model stellar halo   buildup within  the  context of
$\Lambda$CDM   must first   confront the so-called   ``missing satellite
problem'' --- the apparent  over-prediction of low-mass halos compared
to  the abundance of satellite galaxies  around the Milky Way and M31.
For example,  there are eleven know  satellites  of the Milky  Way ---
nine classified  as dwarf spheroidal  and two as dwarf  irregulars ---
yet   numerical work predicts  several   hundred dark matter satellite
halos  in a similar mass  range \citep{klypin99,moore99}.  It is quite
likely that our inventory of stellar satellites  is not complete given
the  luminosity  and   surface brightness  limits   of  prior searches
\citep[as  the  recent discovery of the   Ursa  Minor dwarf spheroidal
demonstrates, see][]{willman05}, but incompleteness is not seen as a viable
solution    for   a  problem   of  this      scale \citep[see][for   a
discussion]{willman04}.

The simplest solution to this problem is to postulate that
only a  small  fraction of   the  satellite halos  orbiting
the Milky Way  host an observable galaxy.
In   this  work,  we   solve the  missing   satellite  problem using the
suggestion of \citet{bullock00},  which maintains that only the  $\sim
10\%$ of low-mass galaxies  ($ \Vmax <  30 \kms$) that had  accreted a
substantial fraction of  their  gas before  the  epoch of reionization
host observable galaxies \citep[see also][]{chiu01,somerville02,benson02,kravtsov04}.
The key  assumption  is that after  the redshift of hydrogen reionization,  $z_{\rm
re}$, gas accretion is suppressed in halos with $\Vmax < 50 \kms$, and
completely stopped in halos with $\Vmax <  30 \kms$.  These thresholds
follow from      the  results     of \citet{thoul96}    and
\citet{gnedin00}  who used hydrodynamic simulations  to  show that gas
accretion in low-mass halos is indeed suppressed in the presence of an
ionizing background. 

We also impose a low-mass cutoff for tracking galaxy
formation in satellite halos with $\Vmax < 15 \kms$.
Two processes and one practical consideration motivate us
to ignore galaxy formation in these tiny halos: first,
 photo-evaporation acts to eliminate any gas that was accreted
before reionization in halos with $\Vmax \lsim 15 \kms$ 
\citep{barkana99,shaviv03}; second,
the cooling barrier below virial temperatures of $\sim 10^{4}$K
(corresponding to $\Vmax \sim 16 \kms$) prevents any gas that could remain bound to these halos
from cooling and forming stars \citep{kepner97,dw03}; finally,
even if we were to allow star formation in these systems, their
contribution to the stellar halo mass would be negligible.
 Once we are more confident of our inventory of the lowest luminosity 
and lowest surface brightness satellites \citep{willman04}of the Milky Way we should 
be able to confirm these physical arguments with observational constraints.

The epoch of reionization $z_{\rm re}$ determines the numbers of galaxies that
have collapsed in each of the above $\Vmax$ limits, and hence the number of luminous satellites
that will be accreted, whether they disrupt to form the stellar halo or survive to form the
Galaxy's satellite system. We discuss limits on this parameter in \S3.1.1.

\subsubsection{Gas accretion following reionization}
\label{sec:gas}

The virial mass of each satellite, 
$M_{\rm vir}^{\rm sat}$,  at the time of its  accretion, $a_{\rm ac}$,
is set by
our  merger tree initial conditions  (\S 2.1.1).  We  assume that each
satellite  halo  has  had a   mass  accumulation  history set 
by Equation \ref{growth} up to the time of
its merger into the  ``Milky  Way'' host, with
$a_0 = a_{\rm ac}$.
 After accretion, all  mass
accumulation onto the satellite is truncated (see \S2.3.3).   
 For massive satellites,
$\Vmax  > 50 \kms$,  we set $a_c$   in Equation \ref{growth} using the
satellite's  mass-defined concentration parameter via Equation
\ref{cofz} \citep[see][]{b01,wechsler02}. This provides a
``typical'' formation   history  for  each satellite.    For  low-mass
satellites, we  are necessarily  interested in where $a_c$ falls  in the
distribution  of halo  formation  epochs  because  this determines the
fraction of mass in place at reionization.  Therefore, if
$\Vmax < 50 \kms$, we use the methods of
\citet{lc93} in order to derive the fraction of the satellite's mass
that was in place at the
epoch of reionization, $z_{\rm re}$, 
and use this to set the value of $a_c$.  
Given $a_c$ for each satellite, we determine the instantaneous
accretion rate of dark matter $h(t)$ in to this system as a function
of cosmic time via
\beq
h(t) =  \frac{\mathrm{d} \Mvir^{\rm sat}}{{\mathrm d}t}.
\eeq

In  the  absence of  radiative feedback  effects, cooling is extremely
efficient in pre-merged satellites of  the size we  consider 
\citep[see,  e.g.][]{mb04}.  Therefore we expect
the cold  gas inflow rate to track
the   dark   matter accretion rate, $h(t)$ --- at least in the 
absence of the effects of reionization --- and take it to be
\beq
 C f_{\rm gas} \, h(t- t_{\rm in}).
 \eeq
The  time  lag
within $h(t)$ accounts for the finite  time it takes for
gas to settle  into the center of the satellite after  being accreted.  
We  assume this occurs in roughly  a halo orbital time at the
virial radius: $t_{\rm in} = \pi R_{\rm h}/\Vvir \simeq 6 \, {\rm Gyr} \,
\,  (1+z)^{-3/2}$.  We have introduced the constant $C$
in order to account for the suppression of gas accretion in low-mass halos
(as alluded to in \S \ref{sec:reion}).
Before the epoch of reionization, we set $C=1$ for all galaxies.
For systems with $\Vmax > 50 \kms$, $C=1$ at all times.  
After reionization, $C=0$  in systems with $\Vmax < 30 \kms$,
and $C$ varies linearly in $\Vmax$ between $0$ and $1$ 
if $\Vmax$ falls between $30$ and $50 \kms$ \citep[see][]{thoul96}.

The fraction of mass in each satellite
in the form of cold, accreting baryons, $f_{\rm gas}$, determines
the total stellar mass plus cold gas mass associated with each dark matter halo.
In what follows,  we adopt $f_{\rm gas} =  0.02$, which is an upper limit on the
range of cold baryonic mass fraction in observed galaxies (Bell et al. 2003).

\subsubsection{Star Formation}
\label{sec:stars}

\begin{figure}
\plotfiddle{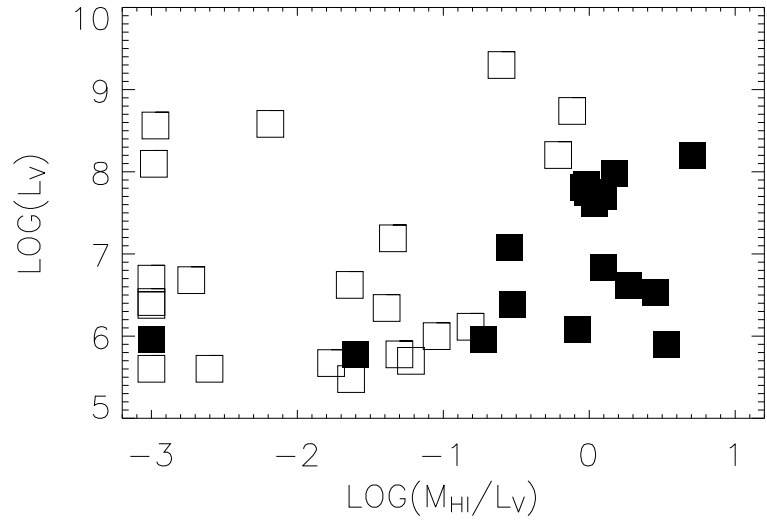}{3.5in}{0}{80}{80}{-170}{0}
\caption{V-band luminosity plotted against the ratio of mass in HI to V-band luminosity for satellites of the Milky Way and Andromeda  (open symbols) and field dwarfs (solid symbols). Data is taken from the compilation by \citet{grebel03}
\label{gasfrac_fig}}
\end{figure}

If we assume that cold gas forms stars over a
timescale  $t_\star$, then the evolution  of  stellar  mass $M_\star$ and cold gas mass $M_{\rm gas}$
follows a simple set of equations:
\beqa
\frac{ \mathrm{d} M_{\star}}{ \mathrm{d}t} & = & \frac{M_{\rm gas}}{t_{\star}}, \label{sfreqa} \\
\frac{\mathrm{d}M_{\rm gas}}{\mathrm{d}t} & = & - \frac{\mathrm{d}M_{\star}}{\mathrm{d}t} + C f_{\rm gas} \, h(t- t_{\rm in}).
\label{sfreqb}
\eeqa
For simplicity, the star
formation is   truncated soon after  each satellite halo
is accreted onto the Milky Way host.
Physically, this could result  from gas loss via
ram-pressure stripping from the    background hot gas halo 
\citep{lin_faber83,moore94,blitz_robishaw00,mb04,Mayer05}.  
This model is broadly consistent with observations that demonstrate that
the gas fraction in satellites of the Milky Way and Andromeda is typically far less than 
that in field dwarfs in the Local Group, as illustrated by the 
separation of the open (satellites) and filled (field dwarfs) 
symbols in Figure \ref{gasfrac_fig} \citep[plotting data taken from][]{grebel03}.
Of  course,  this assumption is
over-simplified, but it allows us to
capture in general both the expectations of  the  hierarchical picture
and the observational constraints.  
We note that this is likely  a bad approximation for
massive satellites, whose deep potential wells will tend to resist the
effects of ram-pressure  stripping.  However, we expect that this will have
little impact on our stellar halo predictions, since most of the
stellar halo is formed from satellites that are accreted early and
destroyed soon after.

The star formation timescale, $t_\star$ determines the star to cold gas fraction 
in each satellite upon accretion and, for a given value of $f_{\rm gas}$, 
total stellar luminosity associated with each surviving satellite and the stellar halo.
We discuss limits on this parameter in \S3.1.2.

\subsection{Embedding baryons within the dark matter satellites}
\label{sec:king}

We  model the evolution of a two-component population of stellar matter
and dark matter in each satellite by associating stellar matter
with the more tightly bound material in the halo.
As discussed in \S \ref{sec:dm}, the mass profile of the
satellite is assumed to take the NFW form.  Mass-to-light
ratios for each particle are picked based on the particle
energy in order to produce a realistic stellar profile for
a dwarf galaxy.

A phenomenologically-motivated approximation for the stellar distribution
in dwarf galaxies is the spherically symmetric King profile (King 1962):
\beq
\label{eq:king}
\rho_{\star}(r) = \frac{K}{x^2} \Bigg(\frac{\cos^{-1}(x)}{x} - \sqrt{1-x^2}\Bigg), \quad x \equiv \frac{1 + (r/r_{\rm c})^2}{1 + (r_{\rm t}/r_{\rm c})^2}.
\eeq 
The core radius is $r_{\rm c}$ and $r_{\rm  t}$
is the tidal radius, where $\rho_{\star}(r>r_{\rm t}) = 0$.  The normalization,
$K$, is set by the average density of the satellite, determined by its
mass (\ref{sec:stars}) and size scales (discussed below).

For each satellite, we assume a stellar mass to light ratio of 
$M_\star/L_{\rm V} = 2$, and use
the stellar mass calculated in \S \ref{sec:stars}
in order to assign a median King core radius 
\beq
r_c = 160 {\rm pc} \left( \frac{L_\star}{10^6 L_{\odot}}\right)^{0.19},
\eeq
where throughout $L_{\star}$ is assumed to be the V-band stellar
luminosity.
We allow scatter about the relation using a uniform
logarithmic deviate between $-0.3 \le \Delta \log_{10} L \le 0.3$.
This slope and normalization was determined by least-square
fit to the  luminosity and core size correlation 
for the dwarf satellite data presented in Mateo (1998), and the
scatter was determined by a ``by-eye'' comparison to the
scatter in the data about the relation.  
Our adopted relation between $r_c$ and $L_\star$ is
also consistent
with the relevant projection of the fundamental plane 
for dwarf galaxies (e.g. Kormendy 1985).
For all satellites we adopt
$r_{\rm t}/r_{\rm c} = 10.$

Assuming isotropic orbits for the stars and that the
gravitational potential is completely dominated by the dark matter, 
the stellar energy distribution function 
corresponding to the King profile $f_\star(\epsilon)$
is determined by setting $\rho=\rho_{\star}$ and $\Psi = -\Phi_{_{\rm NFW}} + \Phi_0$
in equation (\ref{fofe_eq}). 
The mass-to-light ratio of a particle of energy $\epsilon$ is then simply $f_\star(\epsilon)/f(\epsilon)=
(dM/d\epsilon)_\star/(dM/d\epsilon)$.
 Three examples are given Figure \ref{embed_fig}.

\subsection{Limitations of our method}

The main limitation of  our method is  that it only follows the smooth
growth   of    the      parent potential      analytically   ---   the
satellite/satellite interactions and  reaction of  the parent  to  the
satellite  are  not   modeled  self-consistently.   Hence  we   do not
anticipate following the evolution of the field or satellite particles
during a major  or even minor merger  event with great accuracy. Given
this  limitation,  we only  simulate the  accretion histories of halos
generated  from  the Monte Carlo merger tree 
 code  that  have not  suffered a
significant merger  ($>$10 \% of the parent  halo mass)  in the recent
past ($<$7 Gyr)  --- 11 of the  20 accretion histories  generated met
this criterion. In  addition, we consider  results from simulations of
accretion events   that have occurred prior    to the last significant
merger to    be less reliable. We label the halos used in this
work 1-11.  The five   left-hand columns of Table
\ref{halo_tab}   summarize the properties  of  the simulations run for
each halo.

Even  with these restrictions, we  consider our study  to  be a useful
approach for  exploring substructure in  galaxy halos because: (i) the
highest surface  brightness features in halos are  likely to have come
from recent events, whose  debris has had  a shorter time to phase-mix
and/or be dispersed by oscillations   in the parent galaxy  potential;
and (ii) substructure should be more  readily detectable around spiral
(rather than elliptical) galaxies  because their stellar distributions
are less extended --- the existence  of disks in spirals suggests that
these are the ones the more quiescent accretion histories.

\section{Results I: Tests of the Model}

As outlined in \S 2.5, our  method most accurately follows the
phase-space evolution of debris from
accretion   events that occur during relatively   quiescent times in a
galaxy's history (which   we define as  being  after the last  $>$10\%
merger event). In future work we  concentrate on those events. In
this  paper, we analyze   the  results from  simulations  of the  full
accretion histories  of our halos.   While  not accurate in following the
phase-space properties of debris  material from events  occurring
before the epoch of major merging, the fact that these systems
{\it are} disrupted is predicted robustly, and we are
able to record the time of  disruption and   
the  cumulative mass  in those  disrupted events as well.

In what follows, we first constrain the remaining free parameters $z_{\rm re}$ (\S3.1.1)
and  $t_\star$ (\S 3.1.2) (with $f_{\rm gas}=0.02$), by requiring that the general properties of 
our surviving satellite populations 
are consistent with those of the
Milky Way's own satellites. We then go on to
demonstrate that these parameter choices naturally produce
the observed 
distributions in and correlations of
the structural parameters of surviving satellites (see \S3.2.1 and 3.2.2),
as well as stellar halos with total luminosity and radial profiles consistent with
the Milky Way (\S 3.2.3).  

\subsection{Primary constraints on parameters}

\subsubsection{Satellite number counts}

As described in \S2.3.1, we have chosen to solve the missing satellite
problem by suppressing gas accretion in small halos after the epoch of
reionization, $z_{\rm    re}$, and  suppressing  gas  accumulation all
together in satellites smaller   than   $15  \kms$.  The  number    of
satellites that host  stars is then set by  choosing $z_{\rm re}$.  In
the work presented in this paper, we assume that reionization occurred
at a redshift $z_{\rm re} = 10$ or at a lookback time of $13$Gyr.  The
fifth column of  Table  \ref{halo_tab} gives the number  of  luminous
satellites accreted over    the lifetime of each  halo   and the sixth
column gives the number of luminous satellites that survive disruption
in each. (The numbers in brackets are for those events since
the last $>$10\% merger.) We see that  our reionization prescription leads to agreement
within a factor  of $\sim 2$  with  the number of  satellites observed
orbiting the Milky Way.   Our results are  roughly insensitive to this
choice as long as $8 \lsim z_{\rm re}
\lsim 15$.

\subsubsection{Infalling satellite gas content}

When reviewing  the properties of  Local  Group  dwarf galaxies it  is
striking that  --- with the notable exceptions  of the Large and Small
Magellanic  Clouds (hereafter LMC and SMC) ---  satellites of  the Milky  Way  and  Andromeda
galaxies are exceedingly gas-poor compared to their field counterparts
\citep{mateo98,grebel03}. Figure  \ref{gasfrac_fig}    emphasizes this
point by plotting the V-band luminosity {\it vs} gas fraction from the
compilation by   \citet{grebel03} for  satellites  (open  squares) and
field dwarfs (filled  squares). We see that field  dwarfs tend to have
$M_{HI}/L_V \simeq 0.3    -  3$, whereas  satellite  dwarfs   have gas
fractions $\sim 0.001 - 0.1$.  

\begin{figure}
\plotfiddle{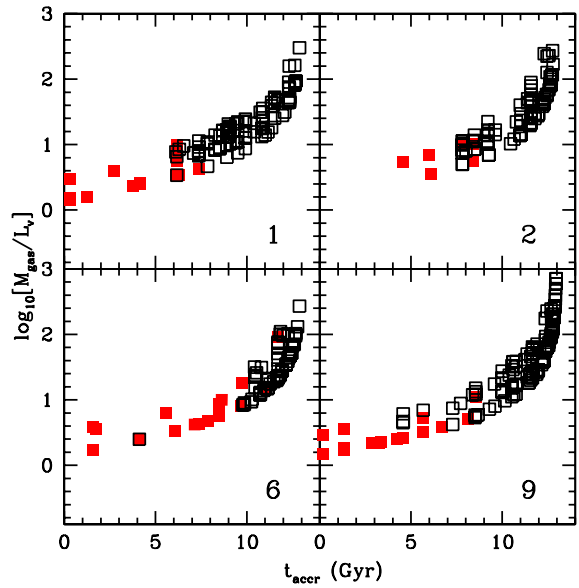}{3.5in}{0}{40}{40}{-150}{-50}
\caption{Plotted is the gas mass per stellar luminosity ratio  $M_{\rm gas}/L_{\rm V}$ 
for each accreted satellite
at the time of their accretion,  $t_{\rm accr}$, for four example halos.
Early accreted systems are more gas rich owing to the rather long star
formation  timescales  in   these  systems.   Solid points
points  indicate
satellites  that  survive  until the  present  day.  
\label{gas*_fig}}
\end{figure}

While  our star  formation  model assumes   that most  of the gas   in
accreted   dwarfs is lost shortly  after  a dwarfs becomes a satellite
galaxy,
consistency with the  field dwarf
population requires that  the  most recent events in  our  simulations
have gas-to-star   ratios of order  unity  immediately prior  to their
accretion.   This  requirement forces us to   choose a  long
star formation
timescale, $t_\star=15$ Gyr, comparable  to the  Hubble time.  Figure
\ref{gas*_fig} shows  the ratio $M_{\rm gas}/L_{\rm V}$
each satellite at the time it was accreted for our four example
halos. 
The
clear trend with accretion time follows because early accreted systems
have not had time to turn their gas into  stars.  Solid points indicate
satellites that survive until  the present day.   We see that the most
recently  accreted systems  ($t_{\rm accr} \sim 1-2$ Gyr,
those that should correspond most  closely
with true  ``field''  dwarfs  today) have 
$M_{\rm  gas}/L_\star  \sim 1 -  2$,
 which  is in  reasonable
agreement with the gas content of field dwarfs.
The points along the lower edge of the trend
 have lower gas fractions at a  fixed
accretion time because they stopped accreting gas at reionization (see
\S2.3.1).


Our choice  of $t_\star = 15$Gyr  is much longer than is
typical for semi-analytic prescriptions of galaxy formation set within
the CDM context
\citep[e.g.,][]{somerville_primack99}, but these usually
focus on much   larger galaxies than the dwarfs we focus on here, where star formation
is likely to have proceeded more efficiently. 
Observations suggest
that  the dwarf  spheroidal satellites  of the Milky  Way  have rather
bursty, sporadic star formation  histories, with recent star formation in some
cases \citep{grebel00,smecker99,
gallart99}.  This effectively demands that the  star
formation timescales must  be long in these  systems: our 
model can be viewed as smoothing over these histories with an
average low-level of star formation.


Note that we do not explicitly  include 
supernova feedback in our star formation histories, but it is
implicitly included by requiring a very low level of efficiency in our
model (i.e. a  large value  of  $t_\star$).
In  two companion papers we do
include the  effects of feedback 
(accounting for both gas gained due to mass loss from stars during normal stellar evolutionary phases
and gas lost via winds driven by supernovae)
in order to
accurately model chemical enrichment in our accreted satellites
\citep{robertson05,font05}.  With feedback included, a choice of
$t_\star   =   6.75$Gyr provides   nearly   identical distributions
of gas and stellar mass in satellite galaxies
as does our non-feedback choice of $t_\star = 15$Gyr.  

\subsection{Verification of Model's Validity}

\subsubsection{Distributions in satellite structural parameters}

\begin{figure}
\plotfiddle{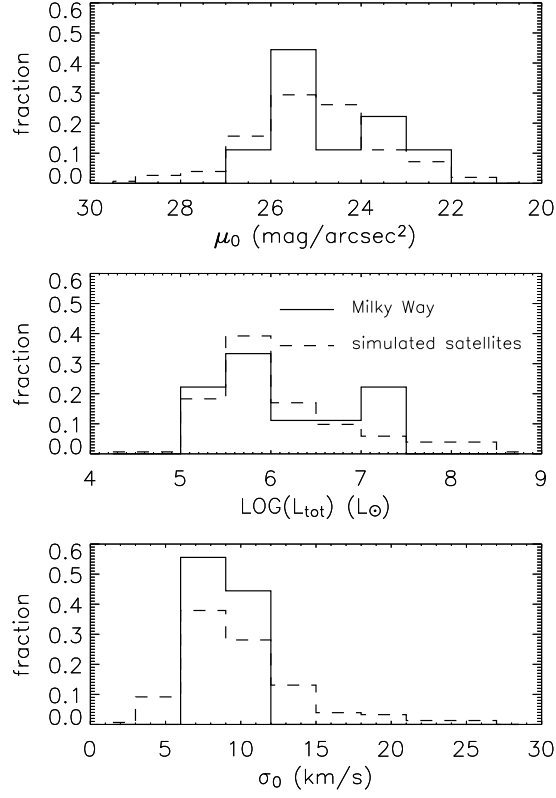}{5.in}{0}{80}{80}{-130}{0}
\caption{Histograms of the distribution of Milky Way 
satellites (solid lines) and simulated satellites (dashed 
lines --- from all 11 halos) as functions of observed quantities 
\label{sat_fig}}
\end{figure}

Figure  \ref{sat_fig} shows  histograms  of  the fractional  number of
satellites as a function of  central surface brightness $\mu_0$, total
luminosity $L_\star$ and central line-of-sight velocity dispersion
$\sigma_\star$ for the Milky Way dwarf spheroidal satellites in solid
lines.
The
dashed lines   represent  our   simulated   distribution of
surviving
satellite properties, derived  by combining  the structural properties
of  the  156 surviving 
satellites  from all   eleven  halos. The histograms  are
visually similar.
(Note that the LMC and SMC are not included in the observational data set
since they are rotationally supported and our models are restricted to hot systems.
They would be equivalent to the most luminous, highest velocity dispersion systems
in our model data set that appear to be missing from the Milky Way distribution.)

\begin{figure}
\plotfiddle{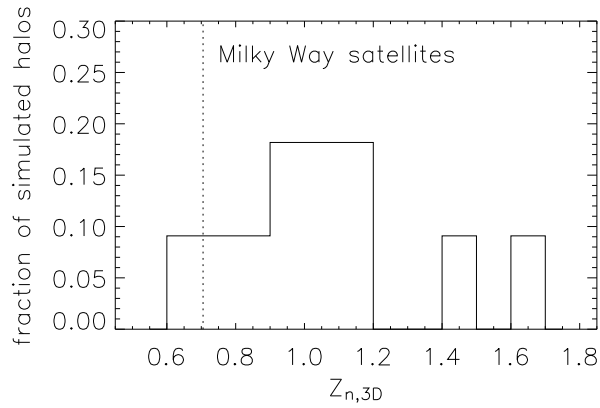}{3.in}{0}{60}{60}{-130}{0}
\caption{Derived KS statistic for the satellite distribution 
in each simulated halo compared with those from the 
combined sample (solid line) and the observed Milky Way 
distribution (dotted line).
\label{kstest_fig}}
\end{figure}

To quantify the level of similarity of the simulated and observed data sets we use the 3-dimensional KS-statistic \citep{fasano87}
\begin{equation}
	Z_{n, 3\rm D}=d_{\rm max}\sqrt{n},
\end{equation}
where $n$ is the number in the  sample tested against our model parent
distribution of all 156 surviving satellites. In this method  $d_{\rm max}$ is  defined as the maximum
difference between  the   observed and predicted  normalized  integral
distributions,  cumulated   within    the  eight  volumes     of   the
three-dimensional space defined for each data point $(X_i,Y_i,Z_i)=(\mu_{0,i},L{\star,i},\sigma_{\star,i})$ by
\begin{equation}
	(x<X_i,y<Y_i,z<Z_i), ..., (x>X_i,y>Y_i,z>Z_i).
\end{equation}
\citet{fasano87} 
present assessments  of the significance level  of values obtained for
$Z_{n, 3\rm D}$ as a function of $n$  and of the degree of correlation
of the data. Since   we  already have eleven  similarly-sized  samples
drawn  from   the same parent  distribution,  we  instead quantify the
significance level  of $Z_{n, 3\rm     D}$ found for the  Milky    Way
satellites by comparing it to the distribution  of $Z_{n, 3\rm D}$ for
our simulated samples.  Figure  \ref{kstest_fig} shows a histogram  of
the results for  our simulated halos,  with the dotted line indicating
where  the Milky Way satellite distribution  falls.  According to this
test only  one of the eleven  simulated populations is more similar to
the simulated parent population  than   the  observed satellites.  (Note    that
$\sim$80\% of our simulated samples  have  $Z_{n, 3\rm D}<1.2$.   This
significance level is similar to those derived by \citet{fasano87} for
3-dimensional samples with $n=10$ and a moderate degree of correlation
in the distribution  --- see their Figure  7 --- as  might be expected
given the expected relation between $\sigma_0$ and $L_{\rm tot}$, see \S3.2.2.)

\subsubsection{Correlations in satellite structural parameters}
\begin{figure}
\plotfiddle{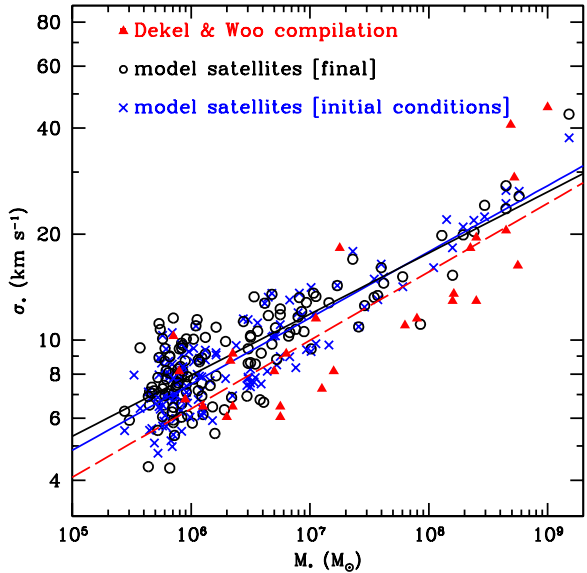}{3.5in}{0}{40}{40}{-150}{-50}
\caption{The relationship between the central, 1-D {\em stellar} 
velocity dispersion and satellite mass for all of our surviving
model satellites.  
Crosses show surviving satellites at the current epoch
and open circles show the relationship computed before the  satellites
were accreted into the host dark matter halo.
Solid triangles show the same relationship for local group satellites
as compiled by \citet{dw03}.  The two nearly identical 
solid lines show the best-fit regressions for the initial and final
model populations.  The dashed line shows the best-fit line
for the data.  The best-fit lines all have similar slopes,
$\sigma_* \propto M_*^{a}$ with $a \simeq 0.2$
($a = 0.19, 0.18$ and $0.19$ for initial, final, and data populations
respectively, and with errors of $0.01$ in $a$ and $\sim 0.07$
in logarithmic normalization.).  The model satellites match
the observed trend in the data quite well, considering the 
observational uncertainties \citep[see, e.g. the discussion in][]{dw03}.
\label{dw_fig}}
\end{figure}

Figure \ref{dw_fig}  shows  the  relationship between  the central
($< r_c$), 1-D  light-weighted
velocity  dispersion and satellite   stellar  mass, $M_\star$,   for
model
galaxies and   observed galaxies  in  the Local   Group.  Crosses show
surviving model satellites for  all halos and  open circles  show the
relationship for the same set  of satellites {\it before they were accreted
into  the  host  dark  matter  halo}.  Solid  triangles  show  the 
relationship for Local Group  satellites as compiled by  \citet{dw03}.
The two nearly identical solid lines show the best-fit regressions for
the  initial and final model populations.    The dashed line shows the
best-fit line for the data.  Our model galaxies  reproduce
a trend quite similar to that seen  in the data.  The relative
agreement   is significant for  two reasons.  First,
the stellar velocity dispersion of  our initialized satellites is  set
by the underlying potential well of their  dark matter halos convolved
with their associated King profile parameters.  While in \S \ref{sec:king}  we
set King profile parameters using a phenomenological relation
based on the stellar luminosity ($L_\star \Rightarrow r_c$), 
there was no guarantee that the dark
matter potential associated with a given luminosity would provide a
consistent stellar velocity dispersion
($r_c + \rho_{_{DM}} \Rightarrow \sigma_\star$).  
In this sense, the general agreement between
model satellites and the data is
a success of our
star formation prescription, which varies
based on the mass accretion histories of halos of a given size (and
therefore density
structure).  

\begin{figure}
\plotfiddle{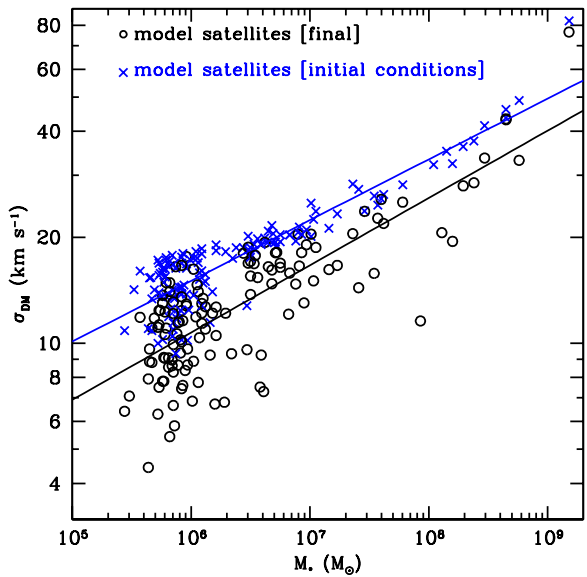}{3.5in}{0}{40}{40}{-150}{-50}
\caption{The  the central, 1-D 
velocity dispersion in the {\em dark matter} versus
 satellite stellar mass.  As in Figure \ref{dw_fig},
crosses show surviving model satellites
and open circles show the relationship computed
before those systems were accreted into the host dark matter halo.
The lines again show least square regression fits,
$\sigma_{\rm DM} \propto M_*^{b}$.  The final systems
tend to have lower velocity dispersions and a broader scatter
at fixed stellar mass owing to the loss of loosely-bound 
dark matter particles after accretion.  Compare this result to
Figure \ref{dw_fig}, where very little shift occurs in the
more tightly bound stellar material.  
\label{vdisp_fig}}
\end{figure}

A  second interesting feature shown in  Figure \ref{dw_fig}  is that {\it final}
surviving satellites obey the same relation as the initial satellites.
Most of these systems have  experienced significant {\it dark  matter}
mass loss, but since the star particles  are more tightly bound, their
velocity dispersion does    not significantly evolve.  This  point  is
emphasized  in Figure  
\ref{vdisp_fig}, where we  plot the  central ($< r_c$),  1-D
velocity dispersion for the {\it dark matter} in halos, again as
a function of the satellite galaxy's stellar mass, $M_\star$.
As in Figure \ref{dw_fig}, open circles show the relationship for the final,
surviving satellites, and crosses show the relationship for those
same satellites before they were accreted.
Unlike in the case of light-weighted dispersions,
the dark matter dispersion velocities in the
surviving systems is systematically lower than in the initial halos
owing to the loss of the most energetic particles.
They also exhibit a broader
scatter at fixed stellar mass, reflecting variations in their
mass-loss histories.
Comparing again to Figure \ref{dw_fig} we see
that most
of the particles associated with light in these systems 
remains bound to the satellites and their velocity dispersions do not
evolve significantly.  This result may have important implications
for interpreting the nature of the dark matter halos of dwarf galaxies
in the Local Group and for understanding the regularity
in observed dwarf properties irrespective of their environments.
 In
future work   we  will  return  to   a  more detailed  structural  and
evolutionary analysis of the light matter and the dark matter halos in
which the  stars are embedded.

The  results presented in this and the previous sub-section 
 clearly   indicate     that our star formation scenario 
coupled with setting   the  
King parameters   of   our infalling  dwarfs     to  match Local  Group
observations  lead to surviving
satellite populations   consistent both in 
number  and structural properties with the Milky Way's.

\subsubsection{The stellar halo's mass and density profile}

Estimates  for the size, shape and  extent of the  Milky Way's stellar
halo    come         either      from  star        count       surveys
\citep{morrison00,chiba00,yanny00,siegel02} or studies where distances
could be estimated using RR Lyraes \citep{wetterer96,ivezic00} . These
studies agree on  a total luminosity of  order $L_{\rm V} \sim
10^9 L_\odot$ (or mass
$\sim 2\times 10^9 \Msun$), which is in good agreement with
the unbound stellar luminosity
for all eleven of our model stellar halos, listed in Column 6 of Table 1
(numbers in brackets again refer to stars from 
accretion events since the last $>$10\% merger).
The match   between  predicted   and   observed total   halo   mass is
non-trivial  and depends sensitively on the  mass accretion history of
the  dark matter  halo  along with the   value  of the  star-formation
timescale, $t_\star$.  Specifically, we show in \S4.1.1 that
 the majority of dwarf galaxies that  make up the stellar
halo were accreted early,  more than $\sim  8$Gyr ago.  
The total stellar halo mass ($\sim 10^{9} \Msun$)
is relatively   small  compared to  the total  cold  baryonic mass  in
accreted satellites ($\sim 10^{10} \Msun$), because the star formation
timescale is long   compared to the   age of the Universe  at  typical
accretion times, and  the stellar  mass fractions are  correspondingly
low  (see Figure \ref{gas*_fig}).  If  we   would have  chosen a  star
formation timescale short compared the time of typical accretion for a
destroyed  system (e.g. $\sim  5$Gyr) this  would  have resulted in  a
stellar  halo of stripped stellar much more  massive than that observed
for the Milky  Way.  
This is in agreement with the results of \citet{brook04b}, who found that a
strong feedback model (effectively slowing the star formation rate in dwarf galaxies) in
their smoothed particle hydrodynamical simulations of galaxy formation
was necessary in order to build  relatively small halo components in their models.

\begin{figure}
\plotfiddle{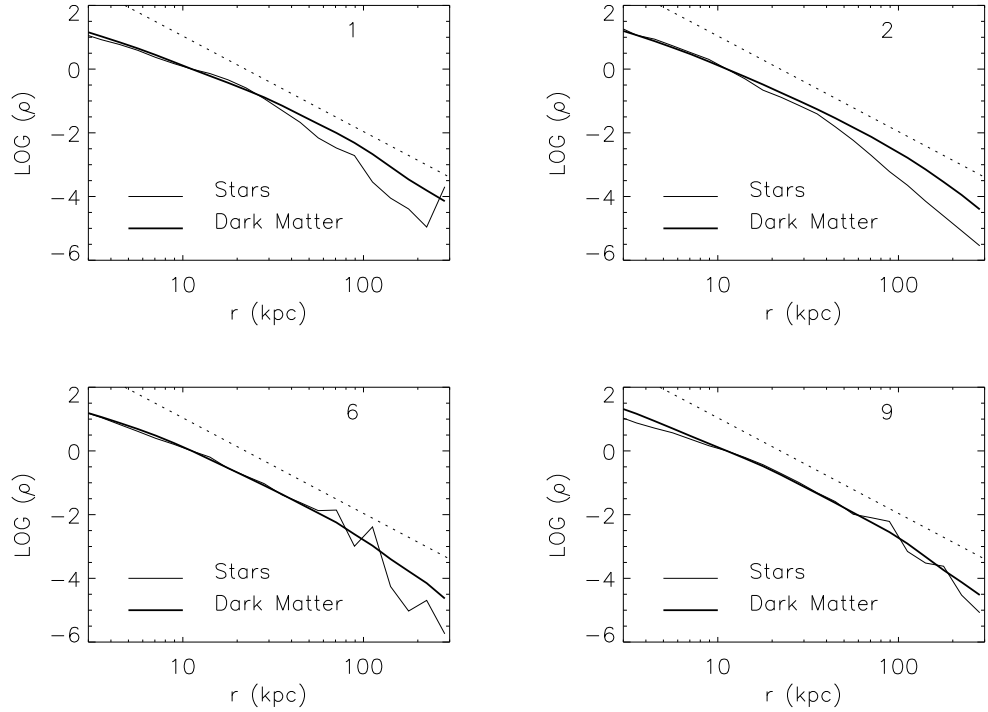}{4.0in}{0}{80}{80}{-190}{0}
\caption{Density profiles for our four example simulated stellar halos (thin solid lines). 
compared to the dark matter halo (bold lines).
Dotted lines represent a power law with exponent -3.
\label{halo_fig}}
\end{figure}

The observational studies find
density profiles falling more steeply than the
dark matter halo (a power law  index in the range  -2.5 to -3.5, compared
to $\sim -2$ for the dark matter at relevant scales). Some
of the variance  between    results  from  different groups    can  be
attributed  to  substructure in  the  halo  since  these  studies have
commonly   been   limited in     sky-coverage with   surveys  covering
significant portions of the sky only now  becoming feasible.  
Figure \ref{halo_fig} plots  the density profiles
generated (arbitrarily normalized)
from    our  four representative stellar  halo  models  
(light solid   curves),  which
transition between slopes of -2 within $\sim 10$kpc to -4 around 50kpc
and fall off even more steeply  beyond this. To illustrate the general
agreement   with observations, the  dotted  line  is  a  power law  with
exponent of -3. Note
that there is some variation in  the total luminosity (about a factor
of 2) and slopes of our model halos, as  might be expected given their
different accretion histories.  There is also  a clear roll-over  below
the power law in the outer parts of the stellar halo, sometimes at 
radii as small as 30kpc.

To contrast to the light,  the density profiles of  the dark matter in
our models  are plotted in  bold lines  Figure
\ref{halo_fig} (also with arbitrary normalization). The  dark
matter profiles  are  all close to  an  NFW profile with
$m_{\rm      halo}=1.4\times  10^{12}      M_\odot$    and     $r_{\rm
halo}=10$kpc. Within $\sim  30$kpc of the  Galactic center it  appears
that our stellar halos roughly track  the dark matter, but beyond this
they tend to fall more steeply. The  difference in profile shapes --- and
the steep roll-over in the light matter at moderate to large radii --- is a natural
consequence of embedding the light matter  deep within the dark matter
satellites: The satellites' orbits  can decay significantly before any
of the more tightly bound  material is lost.  Hence we anticipate that
more/less extended stellar    satellites would result in   a more/less
extended  stellar halo.   Studies  of the  distant  Milky Way halo are
still sufficiently limited that it is not possible  to say whether the
location of the  roll-over in our model stellar  halos  is in agreement
with observations and this could be an  interesting test of our models
in the near future \citep[see, e.g.][]{ivezic03}.

\section{Results II: Model Predictions}

We have now fixed our free parameters to be $z_{\rm re}=10$, $t_\star=15$Gyr 
and $f_{\rm gas}$=0.02.
By limiting our description of the evolution of the baryons associated with
each dark matter satellite to depend on only these
parameters, we find we have little freedom in how we choose them.
For example, if we   were  to
choose a shorter star formation  timescale $t_\star$, we  would over-produce the
mass of the stellar halo, form  dwarf galaxies that were over-luminous
at fixed velocity dispersions, and form  dwarfs with low gas fractions
compared to isolated dwarfs observed in the Local Group.
The first two problems  could be adjusted by adapting  
$f_{\rm gas}$, but the last problem is independent of this.

Despite its simplicity, our 
model reproduces observations of the Milky Way in some detail.
In particular, we recover the full distribution of satellites
in structural properties.
This suggests both that have we assigned the right fraction of dark matter halos
to be luminous {\it and} that our luminous satellites are sitting inside the 
right mass dark matter halos.

We can now go on with some confidence to discuss the implications of our model for
the mass accretion history of the halo and satellite systems (\S4.1)
and the level of substructure in the stellar halo (\S4.2).

\subsection{Building up the stellar halo and satellite systems}

\subsubsection{Accretion times and mass contributions of infalling satellites}

The stellar halo in our model is formed  from stars originally born in
accreted satellites.  Once accreted,  satellites lose mass  with time
until the satellite is  destroyed.  Once a particle  becomes
unbound from a  satellite,  we associate its {\it  stellar} mass
with the stellar  halo.    Figure  \ref{cumfrac_fig} shows  the   cumulative
luminosity fraction of   the stellar halo (solid lines) coming from   accreted satellites as a
function  of  the accretion time   of the   satellite for  halos  1,2, 6, and
9. Clearly most of the mass in the stellar halos originates in satellites
that were accreted more than $8$ Gyr  ago.  
The dotted lines 
 show the contribution to the stellar halo
from  satellite halos more  massive  than  $\Mvir >  2 \times  10^{10}
\Msun$ at the  time  of their  accretion.   
While only
$10-20$  of  the $\sim 150$   accreted  satellites meet  this mass
requirement, we see  that $\sim 75-90\%$  of the mass associated  with
each stellar halo originated within massive satellites of this type.

Compare  these to the dashed  lines,  which show the cumulative number
fraction of  surviving satellite galaxies  as  a function  of the time
they were accreted for  the entire population (long-dashed  lines) and
restricted to satellite halos that were more massive than $\Mvir \gsim
5 \times 10^{9}  \Msun$ at the  time of their  accretion (short dashed
lines).  We  see  that surviving satellites  are  accreted much  later
($\sim 3-5$ Gyr lookback)  than their destroyed counterparts and  that
the most  massive  satellites that survive tend   to be  accreted even
later because the destructive  effects of dynamical friction  are more
important  for   massive   satellites.

\subsubsection{Spatial growth}

Studies of dark matter halos in N-body simulations show that they 
are built  from  the inside  out \citep[e.g.][]{helmi03a}.
The top panel of Figure \ref{rsats_fig} confirms that this idea
holds for our model {\it stellar} halos: it shows the average over all our halos of
the fraction of material in
each spherical  shell from all accretion events (solid line) and
from those  that  have occurred since
the last major ($>10$\%) merger (dotted line --- the time when  this occurred is given
in column 4 of Table  \ref{halo_tab}).
Although the recent events represent
only a fraction of the total halo luminosity ($\sim$5-50\%, see Table
\ref{halo_tab}),   they become the   dominant contributor  at radii of
30-60kpc and beyond. There is some  suggestion  of this  being the case for the
Milky Way's halo globular   cluster population, which can  can  fairly
clearly be separated into an 'old' ,  inner population (which exhibits
some rotation, is slightly flattened  and has a metallicity  gradient)
and a 'young' outer one \citep[which is more extended and has a higher
velocity dispersion --- see][]{zinn93}.

One implication of the inside-out growth of stellar halos, combined with the
late accretion time of surviving satellites is that the two should each follow different
radial  distributions.
The dashed lines in the bottom panel of Figure \ref{rsats_fig}  shows the number
fraction of all surviving satellites in our models as a function of radius --- the distribution is much flatter
than the one shown for the halo in the upper panel.
In fact, all satellites of our own Milky Way (except Sgr) lie at  or beyond  50kpc from
its center, with most 50-150kpc away, --- as shown by the solid line in the lower panel.
Hence, the radial distribution of the observed satellites is consistent with our models
and suggests that they do indeed represent recent accretion events.

\subsubsection{Implications for the abundance distributions of the stellar halo and satellites}

Studies which contrast abundance patterns and stellar populations
in the stellar halo with those in dwarf galaxies 
seem to be at odds with models (such as ours) that build the
stellar halo from  satellite accretion \citep{unavane96}.
For example, both  field and satellite  populations have similar metallicity
ranges, but the former  typically have higher alpha-element abundances
than   the    latter  \citep{tolstoy03,shetrone03,venn04}.
Clearly, it is not possible to build the halo from present-day satellites.

\begin{figure}
\plotfiddle{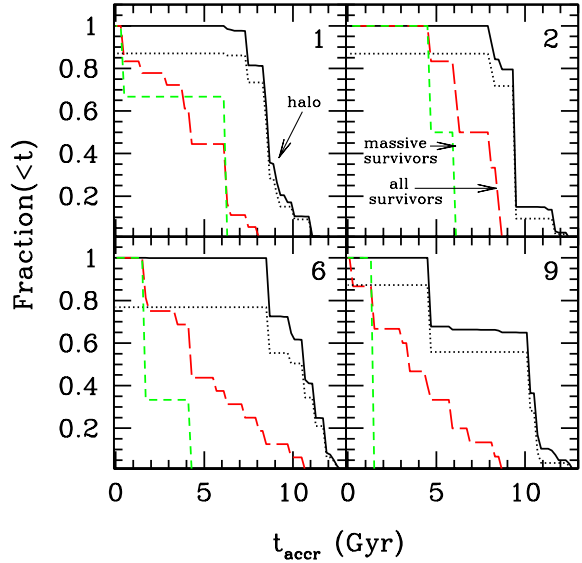}{3.in}{0}{40}{40}{-150}{-50}
\caption{The cumulative 
fraction of stellar halo mass built from accreted satellite galaxies
as a function of the accretion time of the satellites for halos 1-4.
{\em Solid lines} show the full halo and {\em dotted lines} show the contribution from
satellite halos more massive than $\Mvir > 2 \times 10^{10} \Msun$
at the time of their accretion.  While for halos (1,2,6, and 9) only (18,10,12,13)
of the (115,102, 182, 153) accreted luminous satellites were more massive than 
$2 \times 10^{10} \Msun$, we see that $\sim 75-90\%$ of the mass
associated with each stellar halo originated within  massive satellites of this type.
  For comparison, the {\em dashed lines}
show the cumulative fraction of surviving 
satellite galaxies as a function of the time they were accreted.
{\em Short-dashed} lines also show the cumulative accretion times
of surviving satellites, except now
restricted to satellite halos that were more massive than
$\Mvir \gsim 5 \times 10^{9} \Msun$ at the time of their accretion
 We see that surviving
halos tend to be accreted later than destroyed halos. 
There is also a tendency for massive satellites that survive
to be accreted even later because the destructive effects of dynamical friction are more important
for massive satellites.  
\label{cumfrac_fig}}
\end{figure}

\begin{figure}
\plotfiddle{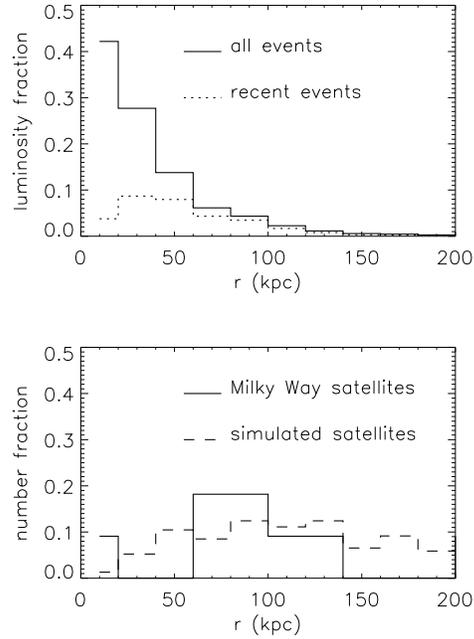}{3.5in}{0}{70}{70}{-150}{0}
\caption{
Top panel compares the luminosity fraction of the entire stellar halo 
spherical shells at radius $r$ from the center of the parent galaxy (solid lines)
with the contribution from events since the last $>$ 10\% merger event (dotted lines).
Bottom panel compares the number fraction of Milky Way satellites in spherical shells as a function
of Galactocentric distance (solid line)
with the number fraction of our surviving satellite population taken from all eleven simulated halos (dashed line).
 \label{rsats_fig}}
\end{figure}

\begin{figure}
\plotfiddle{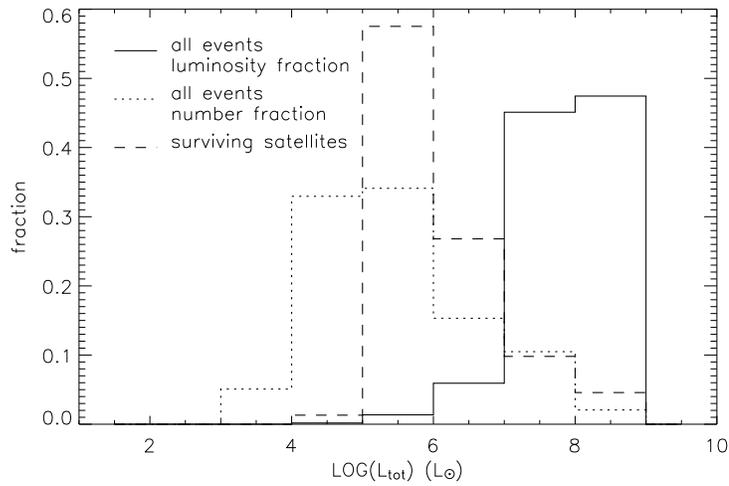}{3.5in}{0}{70}{70}{-170}{-170}
\caption{The number fraction of all events, binned in satellite lunminosity, contributing to all eleven of
our simulated halos (dotted lines) compared to the number fraction of surviving satellites (dashed lines).
The solid lines show the fraction of the total luminosity contributed by each range of satellite luminosities. 
\label{lsats_fig}}
\end{figure}

\begin{figure}
\begin{center}
\plottwo{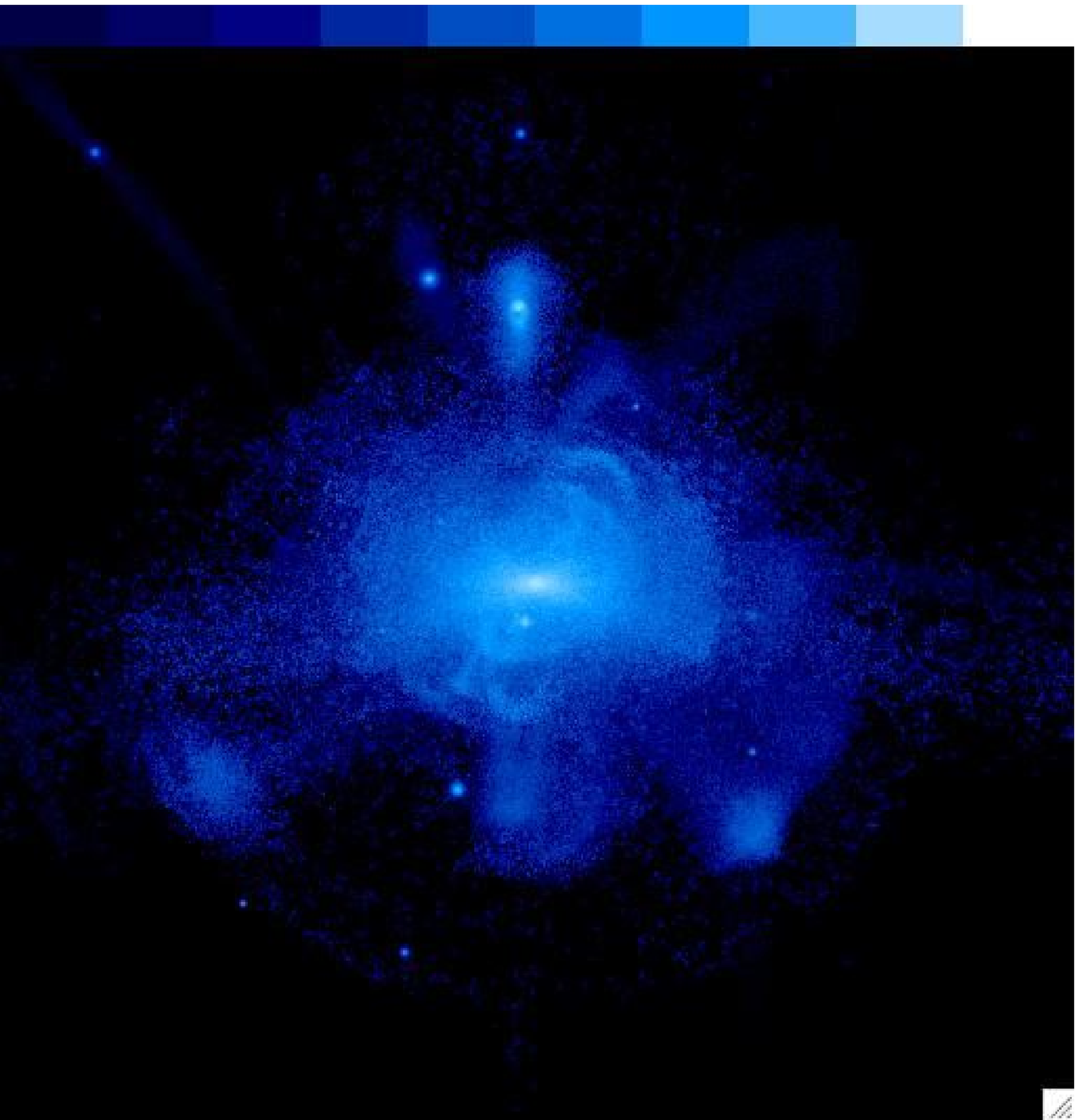}{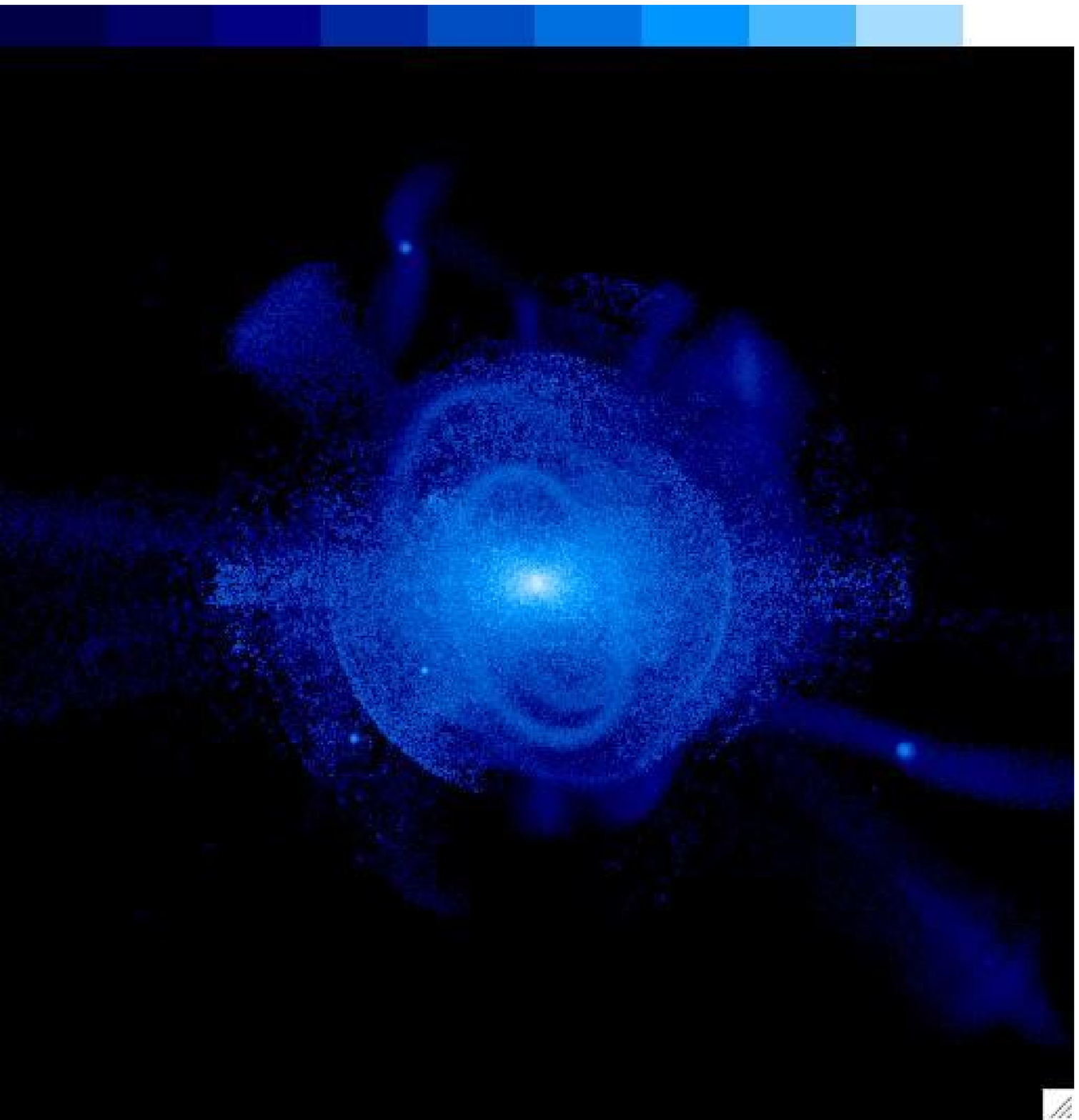}
\caption{``External Galaxy'' views for halo realizations 1 (left) and 2 (right).  The boxes are 300 kpc by 300 kpc.  The blue/white color scale indicates surface brightness: 23 Magnitudes per square arcsecond (white) to 38 Magnitudes per square arcsecond (dark blue / black) where we have assumed a stellar mass to light ratio of 2.   The eye picks up lighter blue (middle of the bar) at about 30 Magnitudes per square arcsecond.
\label{haloviz1_fig}}
\end{center}
\end{figure}

\begin{figure}
\begin{center}
\plottwo{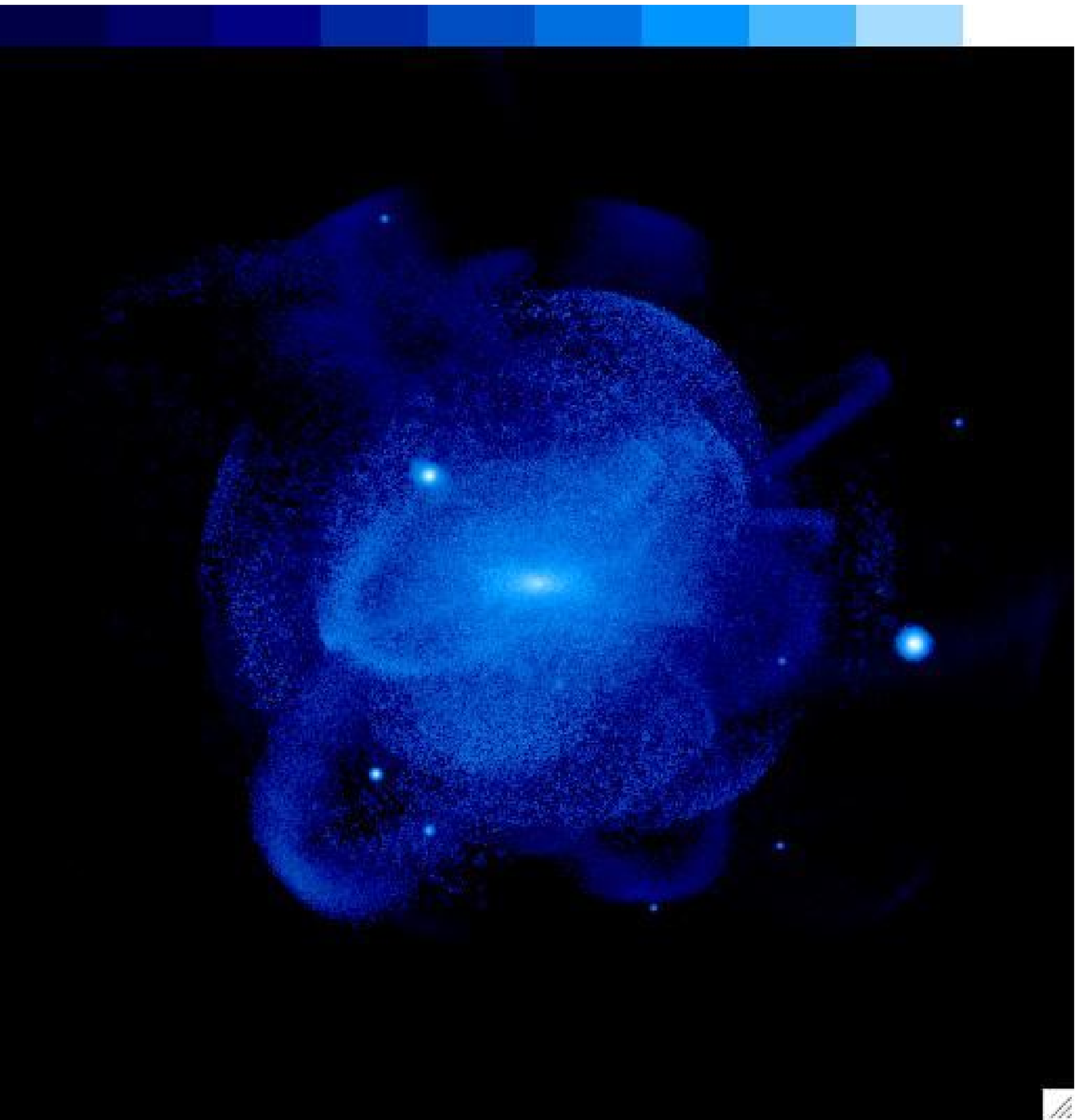}{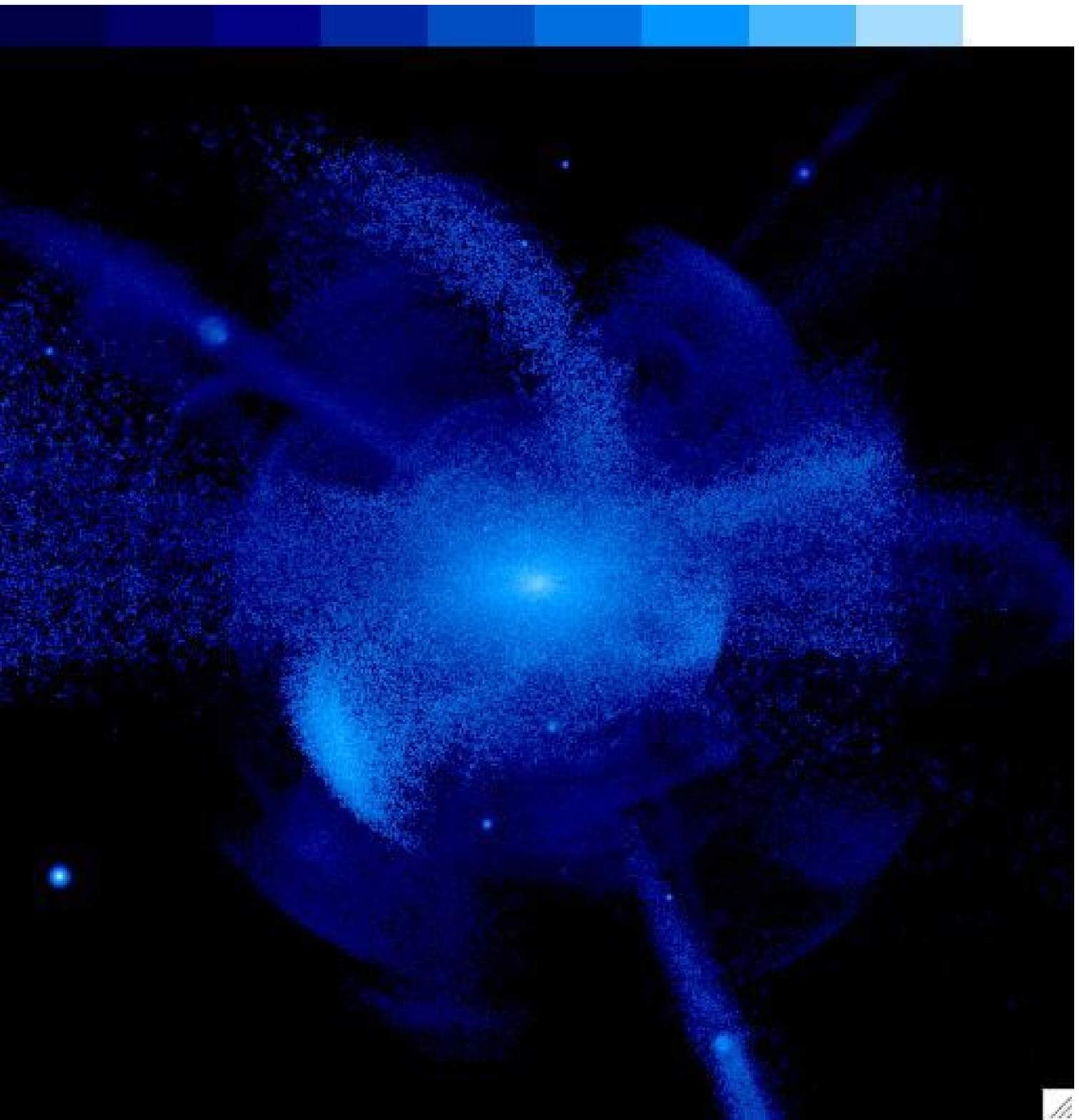}
\caption{``External Galaxy'' images for halos 6 (left) and 9 (right).  The color codes are the same as those in Figure \ref{haloviz1_fig}.
A recent disruption has occurred in halo 9 ($\sim 1.5$Gyr lookback time) and the residue of this event is seen as the
bright plume running from the ``north-west'' of the halo (upper left)
down towards the halo center.  The bright feature just to the ``south-west'' of halo 9's center is also associated
with the same disruption event. 
\label{haloviz2_fig}}
\end{center}
\end{figure}

We have already shown (in Figure \ref{cumfrac_fig} ) that we would expect a random sample of halo stars
to have been accreted 8-10Gyears ago from satellites with masses $\Mvir \gsim 10^{10}\Msun$, while
surviving satellites are accreted much more recently.
(Note that Figure \ref{cumfrac_fig} deliberately compares 
the cumulative {\it luminosity} fraction of the stellar halo to the {\it number} fraction of satellites.
This is the most relevant comparison to make when interpreting observations 
because any sample of halo stars will be weighted by the luminosity of the
contributing satellites, while samples of satellite stars are often composed of
a few stars from each satellite.)
Figure \ref{lsats_fig} explores the number and luminosity contribution of different luminosity satellites to
each population in more detail. It shows the 
number fraction of satellites in different luminosity ranges contributing to the
stellar halo (dotted lines) and satellites system (dashed lines): the peak of the dotted/dashed lines
at lower/higher luminosities is a reflection of the much later accretion time --- and hence longer
time available for growth of the individual contributors --- of the satellite system
relative to the stellar halo.
However, as discussed above, it is more meaningful to compare the number fraction  of
surviving satellites
to the luminosity fraction of the halo (solid lines) contributed by satellites of a given luminosity range.
The solid line emphasizes  (as noted above) that most of the stellar halo
comes from the few most massive (and hence most luminous) satellites, with 
 luminosities in the range $10^7-10^9 L_\odot$. 
In contrast, Galactic
satellite stellar samples
would likely be dominated by stars born in  $10^5-10^7L_\odot$ systems.

Overall our results provide a simple explanation of the difference
between halo and satellite stellar populations and abundance patterns.
The bulk of the stellar halo comes from massive satellites that were accreted early, and hence had star formation histories that
must be short (because of their early disruption) and intense (in order to build a significant luminosity
 in the time before disruption).
In contrast, surviving satellites are lower mass and accreted much later, and hence have more extended, lower level
star formation histories.
Stars formed in these latter environments represent a negligible fraction of the stellar halo in all our models. This is confirmed by the last column of
Table \ref{halo_tab}, which lists the percentage contribution of surviving satellites to the total halo
(less than 10\% in every case).
Note that the contributions of surviving satellites
to the {\it local} halo (i.e. within 10-20kpc of the Sun), 
which is the only region of the halo where detailed abundance studies have been performed,
are even lower (less than 1\% in every case).

A  more quantitative investigation of the
consequences  of the difference between the   "accretion age" of stars
and satellites in the halo is underway \citep{robertson05,font05}.




\subsection{Substructure}

Abundant substructure is  one of  the  most basic  expectations for  a
hierarchically-formed  stellar halo.  Here we give a short description of the
substructure we see in our simulations, 
and reserve more detailed and quantitative explorations for future work.
Recall, our  study (by design) follows  the more recent accretion events
in our halo more accurately than the earlier ones --- we showed in \S4.1.2 that
these are the dominant contributors to the halo at radii of 30-60kpc and beyond.
Hence we can expect our study make fairly
accurate predictions of  expectations of the  level of substructure in
the outer  parts    of  galactic halos   ---
precisely the region where  substructure  should be more dominant  and
easier to detect.

Figures   \ref{haloviz1_fig}  and   \ref{haloviz2_fig} show   external
galactic views for halo  realizations 1, 2, 6, and  9.  The color code
reflects surface brightness per pixel: white,  24 magnitudes per square
arcsecond, to  light blue at  30 magnitudes  per  square arcsecond,  to
black which is (fainter than) 38 magnitudes per square arcsecond.  The
darkest blue  features are of  course too faint to  be seen (except by
star counts).  We have simply set the scale in order to reveal all the
spatial features that are there in principle.  We mention that our test
particles (\S \ref{sec:test}) were used in making these images.

\begin{figure}
\begin{center}
\plottwo{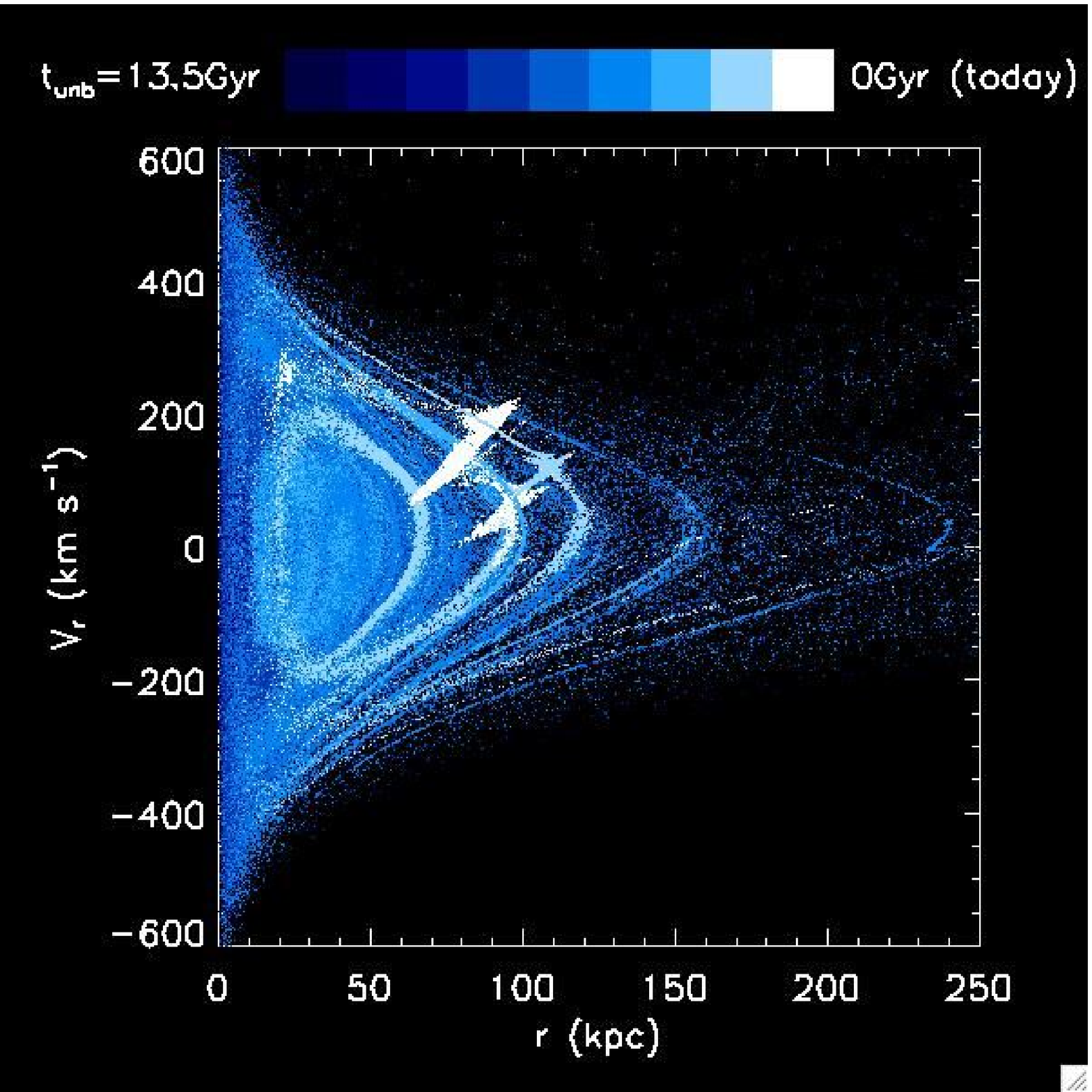}{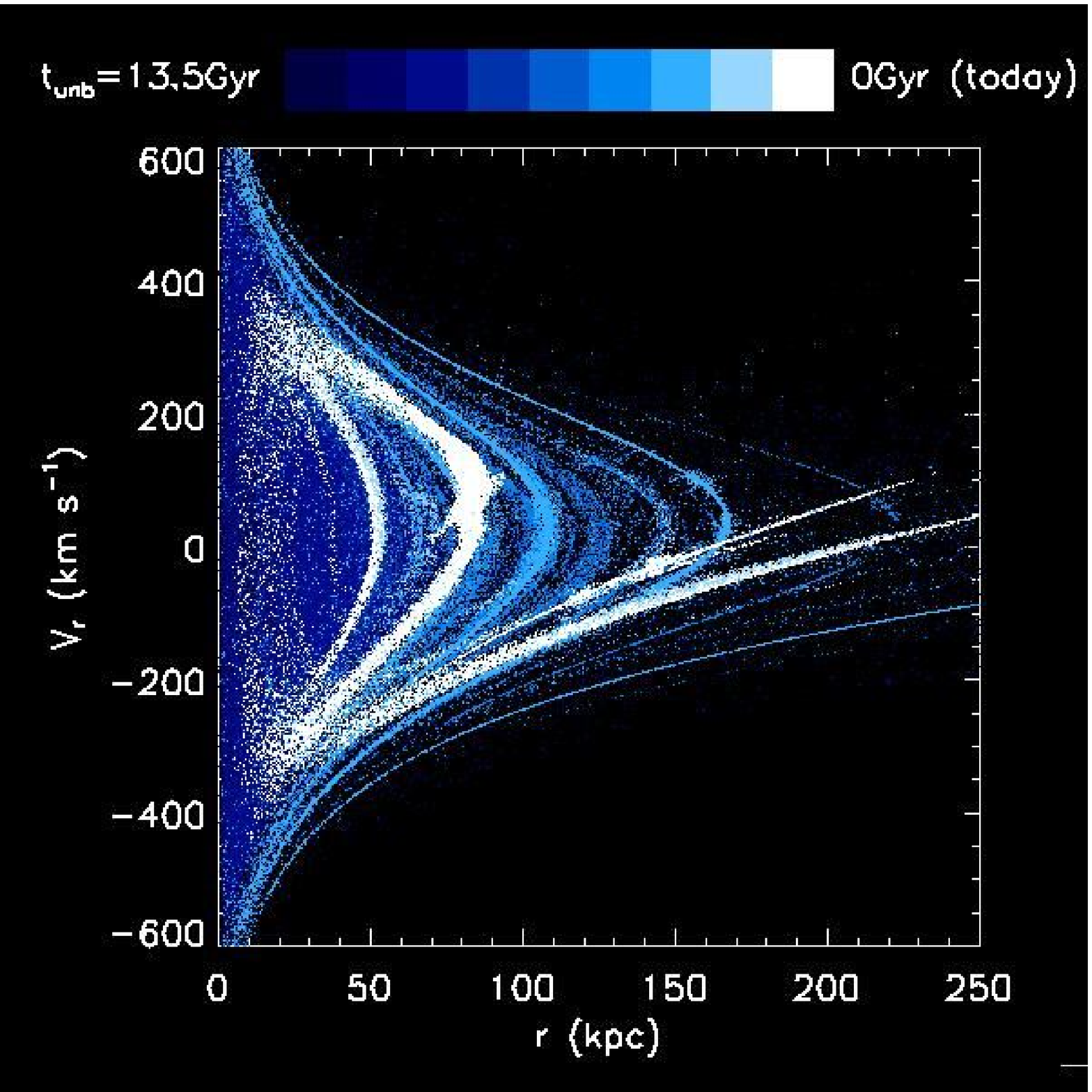}
\caption{Radial phase space diagrams ($V_r$ vs $r$ relative to the host halo center) for  halos 1 (left) and 9 (right).  Each point represents 1000 solar luminosities.  
The  color code reflects the time  each particle became unbound to its
parent satellite.  White points are either bound or became unbound in the
last 1.5 Gyr, while dark blue points became unbound more than 12 Gyr ago.
The radial color gradient reflects the tendency for inner halo stars to be
accreted (and stripped) early in the Galaxy's history.  The white feature at $r \sim 80$ kpc
in halo 9 represents a very recent disruption event -- the most recent massive
disruption seen in our ensemble of 11 halo realizations.
\label{phase_diagram1_fig}}
\end{center}
\end{figure}

\begin{figure}
\begin{center}
\plottwo{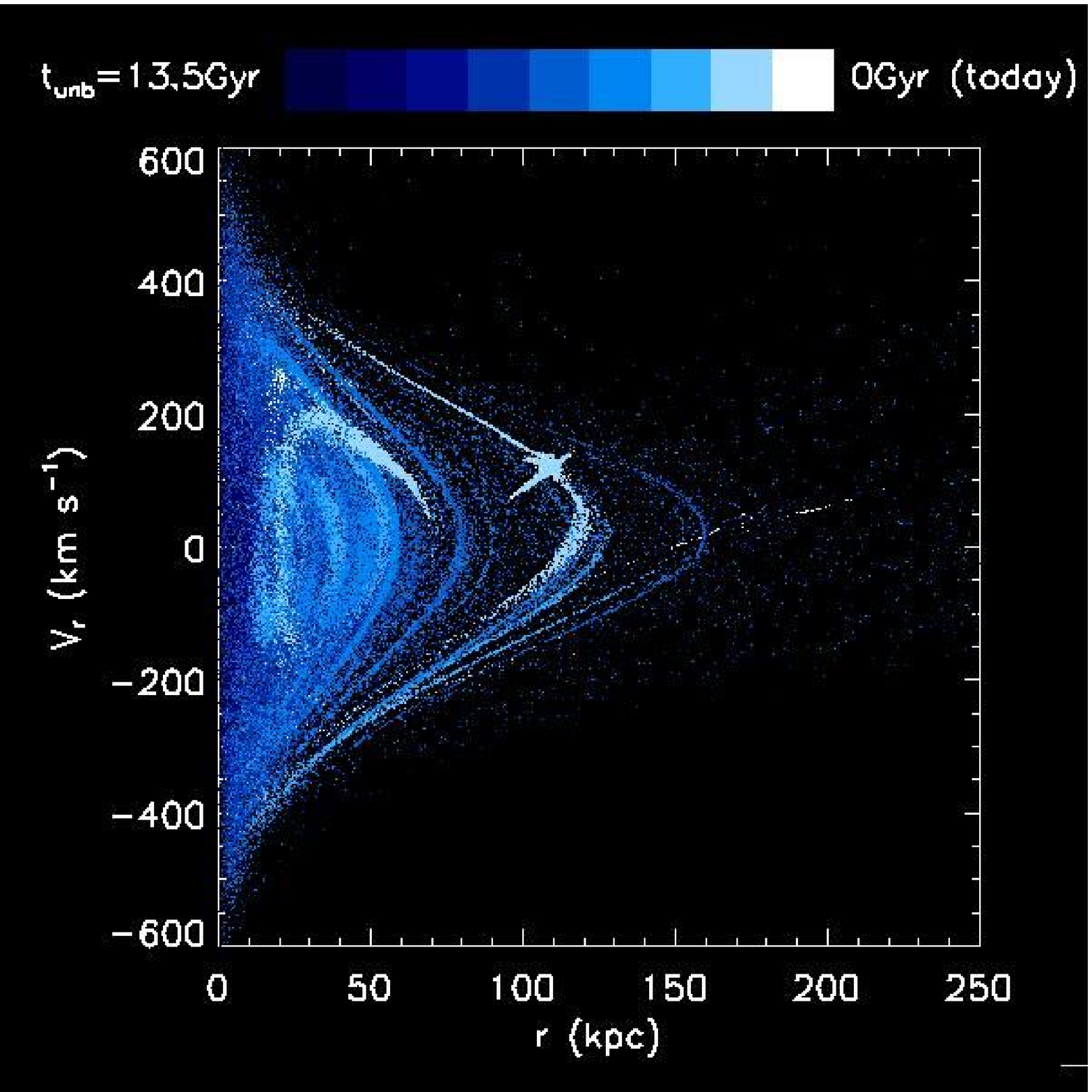}{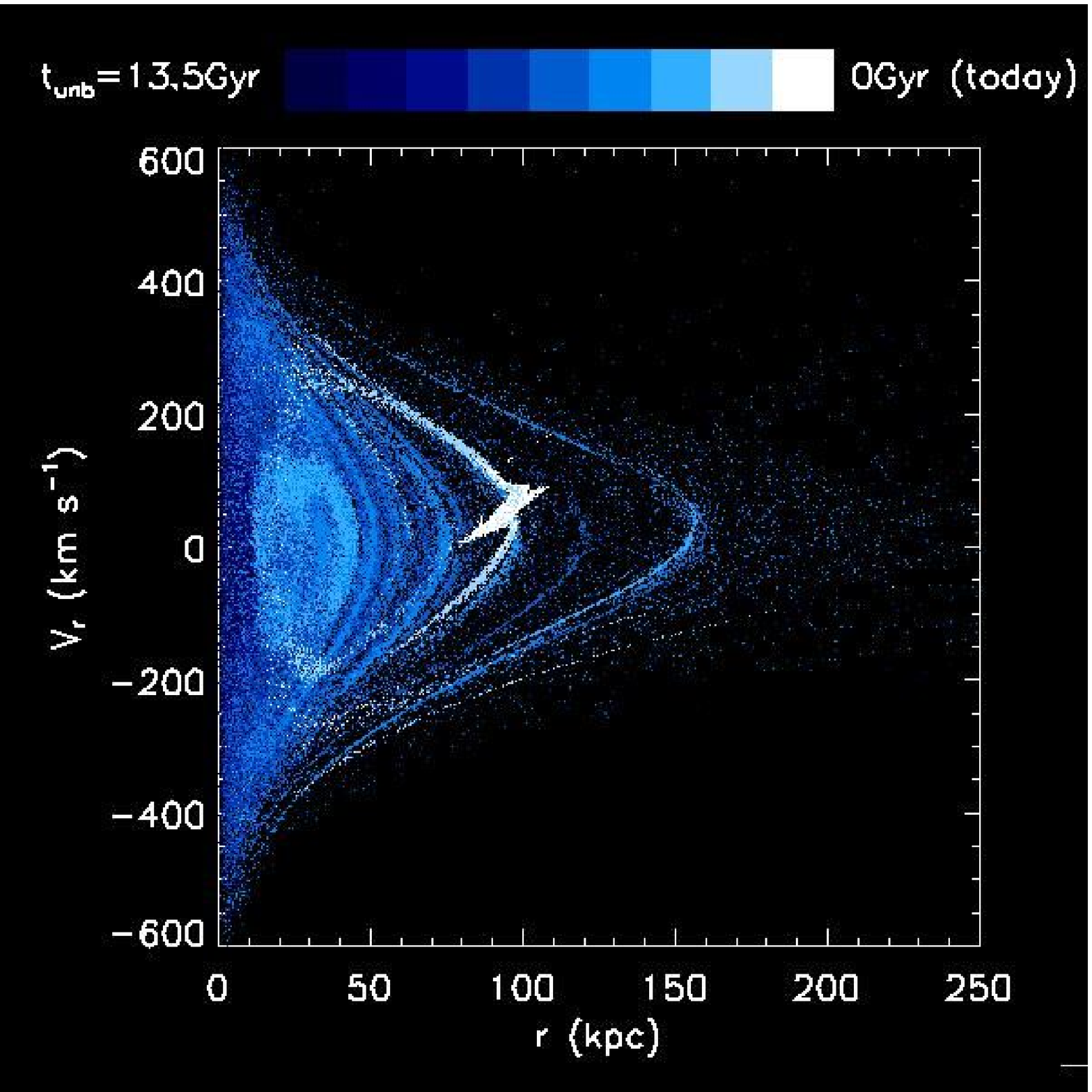}
\caption{Radial phase space diagrams for halo 2, where the left and right figures represent to separate quarters of the sky.  The color code and axis labels are the same as those in Figure \ref{phase_diagram1_fig}.
\label{phase_diagram2_fig}}
\end{center}
\end{figure}

In addition to spatial structure, we also expect significant structure
in phase  space.  A two-dimensional slice of  the full six dimensional
phase space  is illustrated in  Figure \ref{phase_diagram1_fig}, where
we plot radial velocity $V_r$ versus radius $r$  for all
of halo 1 (left) and
halo  9    (right).    Each      point   represents  1000        solar
luminosities.~\footnote{ In   most cases,  we subsample   our luminous
particles in  order  to plot a   single point for every  $1000 \Lsun$.
However,   some of the particles   in  our simulations have  luminosity
weights greater than 1000 $\Lsun$.  In these cases we plot a number of
points ($= L_{\rm particle}/1000$) with the same  $V_r$ and $r$ as the
relevant particle (using small random offsets to give  the effect of a
``bigger'' point on the plot).}  The  color code reflects the time the
particle  became  unbound from its  original  satellite: dark blue for
particles that became  unbound more  than 12   Gyr ago and  white  for
particles that either remain bound or became unbound less than 1.5 Gyr
ago.     The  radial gradient  in  color  reflects  the ``inside out''
formation of the stellar halo discussed in previous sections.

Note    that  significant coherent    structure is  visible  in Figure
\ref{phase_diagram1_fig} even without any spatial slicing of the halos. 
 Except for particles belonging to bound satellites (white streaks),
the structure strongly resembles a nested series of orbit diagrams.
This is not surprising since the halo was formed by particles brought
in on satellite orbits.  A direct test of this prediction should be
possible with SDSS data and other similar surveys.  Indeed, if the phase
space structure of stellar halo stars does reveal this kind of
orbit-type structure it will be a direct indication that
the stellar halo formed hierarchically.  Figure \ref{phase_diagram2_fig}
shows the same diagram for halo 1, now subdivided into two distinct
quarters of the sky.

\section{Summary and Conclusions}

We  have presented a   cosmologically  self-consistent model for   the
formation of the  stellar    halo in Milky-Way  type  galaxies.    Our
approach  is hybrid.  We use a  semi-analytic formalism to calculate a
statistical ensemble of  accretion histories for  Milky Way size halos
and to   model star formation   in  each accreted  system.   We use  a
self-consistent N-body approach to follow  the dynamical evolution 
of   the accreted satellite galaxies.   A crucial
ingredient   in our model  is   the explicit  distinction between the
evolution  of light and  dark  matter in  accreted galaxies.   Stellar
material is much  more  tightly bound than   the majority of  the dark
matter in accreted halos and this plays an important role in the final
density distribution  of stripped  stellar material   as  well as  the
evolution in the observable quantities of satellite galaxies.

A primary goal of this, our first paper on stellar halo formation, was
to normalize our model to, and demonstrate consistency with, the gross
properties of  the Milky Way  stellar  halo and  its satellite  galaxy
population.   We constrained our two  main star formation parameters, the
redshift of  reionization    $z_{\rm re}$,  and   the  star  formation
timescale of cold gas $t_*$, using the observed number counts of Milky
Way satellites and the  gas mass fraction  of isolated dwarf galaxies.
With these parameters fixed, the model reproduces many of the observed
structural properties of    (surviving)  Milky Way  satellites:    the
luminosity function,  the luminosity-velocity dispersion relation, and
the surface brightness  distribution.  The satellite galaxies that are
accreted and destroyed in our model produce  stellar halos of material
with a total luminosity in line with estimates for the stellar halo of
the Milky Way ($\sim 10^9 L_{\odot}$).

The success of our model lends support  to the hierarchical stellar
halo formation scenario, where the stellar
halos of large galaxies form mainly via  the accretion subsequent
disruption of smaller galaxies. More specifically, it allows us to
make more confident predictions concerning the precise nature of
stellar halos and their associated satellite systems in Milky Way type galaxies.  These include:

\begin{itemize}

\item The density profile of the stellar halo should follow a 
varying power-law distribution, changing in radial slope from $\sim -2$
within $\sim 20$kpc  to $\sim -4$   beyond 50kpc.  The  distribution is
expected to be much more  centrally concentrated than the dark matter,
owing to the fact that the stars that build the stellar halo were much
more  tightly bound to   their  host systems  than  the dark  material
responsible for building up the dark matter halo.

\item Stellar halos (like dark matter halos) are expected to form from the
inside out, with the majority of mass being deposited from the $\sim 15$ most massive accretion events, typically dwarf-irregular size halos with mass
$\sim 10^{10} \Msun$ and luminosities of order $10^7-10^9 L_\odot$.

\item Destroyed satellites contributing mass to the stellar halo tend to be
accreted earlier than satellites that survive
as present-day dwarf satellites ($\sim 9$Gyr compared to 
$\sim 5$ Gyr in the past).

\item Substructure, visible both spatially and in phase space diagrams, should be abundant in the outer parts of galaxies.  Proper counts 
of this structure,  both  in our  galaxy and  external systems, should
provide important constraints  on the late-time accretion histories of
galaxies and a test of hierarchical structure formation.

\end{itemize}

Together, the second and third points imply that
most  of  the stars  in the  inner  halo  are associated  with massive
satellites that were accreted $\gsim 9$Gyr  ago.  Dwarf satellites, on
the other hand,  tend to be lower mass  and are  associated with later
time accretion events.  This suggests that classic ``stellar halo''
stars should be quite    distinct chemically from stars   in surviving
dwarf satellites.  We explore this point further in two companion
papers (Robertson et al. 05; Font et al. 05).

\acknowledgments
KVJ's contribution was supported through NASA grant NAG5-9064 and NSF CAREER award 
AST-0133617.



\clearpage

\begin{table}[h]

\fontsize{10}{10pt}\selectfont

\begin{tabular}{|c|c|c|c|c||c|c|c|c|c|c|c| }


\hline
       
	& \#		&		& time of  	&  \#	   		& \#   		& stellar halo       	& \% halo 	  	& 80\% halo  	& 80\% halo 		& \% of halo	\\
Halo	& satellites &$a_c$ 	& last $>$10\% 		& satellites	& surviving   	& luminosity	     	& from 15   	& accretion   	& accumulation 	&  from \\
	& in	merger &		&  merger 		& simulated	&  satellites	& ($10^9 L_\odot$)	& largest 		& time   		& time			&  surviving	\\
	& tree 	&  	&	(Gyr)		& 	 		& 	   			&& satellites        & (Gyr)  		& (Gyr) 			& satellites  	\\

\hline
\hline
Milky Way& --- & ---  & 8-10?		& ---			& 11			& $\sim 1$	     & ---		  & ---			& ---			&	---		\\ 
\hline
\hline
1 	 &391 	&0.375	& 8.3			& 115 (57)	& 18	(18)		& 1.2 (0.29)	     &	87 \%		  & 8.4 		& 5.3 			& 0.96\%		\\ 
\hline
2 	 &373	& 0.287 	& 9.2			& 102 (45)	& 6 	(6)		& 1.1 (0.35)	     &	87 \%		  & 8.6   		& 7.0			& 0.03\%	  \\  
\hline
3 	& 322	& 0.388	& 8.9			& 106 (47)	& 16 (15)		& 0.95 (0.05) 	     &	79 \% 		  & 9.0 		& 7.4			& 0.12\%	 \\ 
\hline		
4 	& 347	& 0.393 	& 8.3			& 97 (32)		& 8 (7) 		& 1.33 (0.14)	     &	91 \%		  &  8.3  		& 6.3			& 0.40\%	 \\ 
\hline
5 	& 512	& 0.214 	& 10.8	 	& 160 (115)	 & 18	 (18)		& 0.68 (0.44) 	     &	78 \%		  & 7.0 		& 2.1			& 0.25\%	 \\ 
\hline
6 	& 513	&0.232 	& 10.5		& 169 (68)	& 16 	(15)		&  0.60 (0.24) 	     &	77 \%		  & 8.6			& 6.2			& 0.01\%	 \\ 
\hline
7 	& 361	& 0.385 	& 7.4			& 102 (48)	& 20 (18)		& 0.70 (0.20) 	     & 	82 \%		  & 7.2			& 4.4			& 8.42\%	 \\ 
\hline
8 	& 550	& 0.205 	& 9.3			& 213 (62)	& 13 (13)		& 0.64 (0.201) 	     &	80 \%		  & 8.8			& 7.1			& 2.55\%	 \\ 
\hline	
9 	& 535	& 0.187  	& 10.0		& 182 (63)	& 15 (15)		& 0.85 (0.36) 	     &	87 \%		  & 4.7			& 1.5			& 0.01\%	 \\ 
\hline
10 	& 484	&0.229 	& 9.7			& 156 (76)	& 13 (13)		& 1.02 (0.65) 	     & 	80 \%		  & 6.7			& 2.9			& 0.04\%	 \\ 
\hline
11 	& 512	& 0.230 	& 9.0			&  153 (63)	& 10 (10) 		& 0.84 (0.22)		&	89 \%		  & 9.1			& 7.2			& 0.02\%	 \\ 
\hline

\end{tabular}
\caption{Properties of our simulated stellar halos.} 
\tablecomments{Numbers in brackets in columns 5, 6 and 7 are for events since the last $>$ 10\% merger (the time of this event is given in column 4).}
\label{halo_tab}
\end{table}

\normalsize

\clearpage

\end{document}